\documentclass[11pt,a4paper]{article}
\usepackage{amsmath, amssymb, amsthm, mathrsfs}
\usepackage[a4paper,margin=1in]{geometry}
\usepackage{hyperref}
\usepackage{xcolor}
\usepackage{booktabs}
\hypersetup{
    colorlinks=true,
    linkcolor=blue,
    filecolor=magenta,      
    urlcolor=cyan,
    citecolor=blue,
}

\usepackage{graphicx}
\graphicspath{{./simulation_pic/}}

\numberwithin{equation}{section}

\newtheorem{theorem}{Theorem}[section]
\newtheorem{lemma}[theorem]{Lemma}
\newtheorem{proposition}[theorem]{Proposition}
\newtheorem{corollary}[theorem]{Corollary}
\newtheorem{definition}[theorem]{Definition}
\newtheorem{example}[theorem]{Example}
\newtheorem{assumption}[theorem]{Assumption}
\newtheorem{remark}[theorem]{Remark}

\newcommand{\Nabla}{\overline{\nabla}}
\newcommand{\Div}{\overline{\mathrm{div}}}
\newcommand{\dd}{\mathrm{d}}

\usepackage{bigfoot}
\DeclareNewFootnote{A}[arabic]
\DeclareNewFootnote{B}[fnsymbol]
\usepackage{etoolbox}
\makeatletter
\patchcmd\maketitle{\def\@makefnmark{\rlap{\@textsuperscript{\normalfont\@thefnmark}}}}{}{}{}
\makeatother
\makeatletter
\def\thanksA#1{
  \footnotemarkA\protected@xdef\@thanks{\@thanks%
        \protect\footnotetextA[\the \c@footnoteA]{#1}}%
}
\def\thanksB#1{%
  \footnotemarkB%
  \protected@xdef\@thanks{\@thanks%
        \protect\footnotetextB[\the \c@footnoteB]{#1}}%
}
\makeatother

\AtBeginDocument{%
  \apptocmd{\normalsize}{%
    \setlength{\abovedisplayskip}{6.5pt plus 2pt minus 4pt}%
    \setlength{\belowdisplayskip}{6.5pt plus 2pt minus 4pt}%
    \setlength{\abovedisplayshortskip}{0pt plus 2pt}%
    \setlength{\belowdisplayshortskip}{4pt plus 2pt minus 2pt}%
  }{}{\PackageError{layout}{Cannot adjust display spacing}{}}%
  \normalsize
}

\begin{document}

\title{\textbf{
Covariant Calculus and Hamiltonian Flows for Nonlocal Wasserstein Metrics
}}
\author{%
  {\normalfont\normalsize\bfseries
    Yuanke Chen\thanksA{School of Mathematics, Sun Yat-sen University, Guangzhou 510275, P.R. China
    \&
    School of Science, Great Bay University, Dongguan 523808, P.R. China.
    Email: \texttt{chenyk37@mail2.sysu.edu.cn},
    \texttt{233621001@stu.gbu.edu.cn}.
    },
    Jinqiao Duan\thanksA{%
    School of Science, Great Bay University, Dongguan 523808, P.R. China.
    Email: \texttt{duan@gbu.edu.cn}.
    },
    Qiao Huang\thanksA{School of Mathematics, Southeast University, Nanjing 211189, P.R. China. Email: \texttt{qiao.huang@seu.edu.cn}}
    $^{,}$\thanksB{Corresponding author}%
  }%
}
\date{}
\maketitle

\vspace{-0.5in}

\begin{abstract}
In this paper, we study nonlocal metrics on smooth positive probability densities on a closed manifold, induced by symmetric jump kernels and density-dependent mobilities. First, we introduce the density-variation operator to express the density derivatives of the Dirichlet form and kinetic energy. Second, we construct density and velocity-field representations of the tangent space and prove their isometric equivalence. We use the generator's pseudoinverse to derive Levi-Civita connections and covariant differentials in both representations. Third, we establish mechanical Hamiltonian dynamics in both tangent representations; an action principle and Legendre duality identify the Hamilton--Jacobi and covariant Newton equations; smooth negative gradient flows admit a zero-energy Hamiltonian lift. Entropy diffusion, nonlocal mean-field equations, Keller--Segel dynamics, and a nonlocal Vlasov equation illustrate the resulting calculus. For geometric-mean mobility, the Hopf--Cole transformation of the nonlocal Schr\"odinger bridge system and the Madelung transformation of the nonlocal Schr\"odinger equation yield hyperbolic and trigonometric phase energies, respectively. At zero momentum, both recover the metric kinetic energy to second order.
\end{abstract}




\section{Introduction}

Optimal transport links the geometry of probability measures to variational evolution equations. The dynamic formulation of Benamou and Brenier expresses quadratic Wasserstein transport through a continuity equation and kinetic action \cite{Benamou+Brenier_2000_Computational}. The Fokker--Planck variational scheme and Otto's interpretation of porous-medium dynamics established the gradient-flow viewpoint \cite{Jordan+Kinderlehrer+Otto_1998_Variational,Otto_2001_Geometry}, followed by a general metric theory \cite{Ambrosio+Gigli+Savare_2008_Gradient,Villani_2009_Optimal}. For differential calculations on smooth positive densities, Lott's geometric calculus \cite{Lott_2008_Geometric} complements measure-theoretic second-order calculus \cite{Gigli_2012_Second}. Related approaches use diffeomorphism groups and invariant metrics \cite{Khesin+Lenells+Misiolek+Preston_2013_Geometry,Bauer+Modin_2020_SemiInvariant,Bauer+Bruveris+Michor_2016_Uniqueness,Bauer+Joshi+Modin_2017_Geodesic,Bruveris+Michor_2019_Geometry}, with connections to quantum mechanics and information geometry \cite{Molitor_2015_Relation,Ay_2025_Information}; infinite-dimensional background is provided by \cite{Kriegl+Michor_1997_Convenient}.

For jump dynamics, mass moves between pairs of states. Maas and Mielke constructed transport metrics for reversible finite Markov chains, with logarithmic-mean mobility realizing the Markov evolution as an entropy gradient flow \cite{Maas_2011_Gradient,Mielke_2013_Geodesic}; entropic curvature was studied in \cite{Erbar+Maas_2012_Ricci}. Erbar extended the transport framework to continuous-state jump kernels \cite{Erbar_2014_Gradient}, and Slep\v{c}ev and Warren studied nonlocal distances and their relation to local transport \cite{Slepcev+Warren_2023_Nonlocal}. Further developments concern distribution-dependent chains, graph interaction equations, and stationary point processes \cite{Erbar+Fathi+Laschos+Schlichting_2016_Gradient,Esposito+Patacchini+Schlichting+Slepcev_2021_Nonlocal,Huesmann+Stange_2025_Nonlocal}. Symmetric singular jump kernels also admit a Dirichlet-form treatment \cite{Fukushima+Oshima+Takeda_2011_Dirichlet,Barlow+Bass+Chen+Kassmann_2009_Nonlocal}.

The mobility determines how the metric depends on the density. Local nonlinear mobilities have been studied through modified transport distances and displacement convexity \cite{Dolbeault+Nazaret+Savare_2009_New,Carrillo+Lisini+Savare+Slepcev_2010_Nonlinear,Lisini+Marigonda_2010_Class}. Differential formulas for graph metrics and mean-field information Hessians appear in \cite{Li_2018_Transport,Li+Lu_2022_Mean}. Li derived Levi-Civita connections, parallel transport, and curvature formulas for finite-state probability manifolds and local hydrodynamical density manifolds \cite{Li_2025_Geometric,Li_2026_Geometric}.
In contrast, to the best of our knowledge, Levi-Civita connections
and covariant derivatives for nonlocal transport metrics in continuum have not been systematically developed. Such metrics typically
involve a density-dependent jump kernel weighted by a two-point
mobility $\theta(a,b)$ (see e.g.,
\cite{Erbar_2014_Gradient}):
\begin{equation}\label{eq:intro-weighted-kernel}
J_\rho(x,y)=\theta\bigl(\rho(x),\rho(y)\bigr)K(x,y),
\end{equation}
where $K(x,y)$ is a reference jump kernel.
Differentiating the resulting metric to obtain its Levi-Civita
connection requires tracking both $J_\rho$ and the potentials
representing fixed density tangent vectors. In particular, the nonlocal carr\'e du champ alone does
not capture the covariant derivative in the velocity-field
representation of the tangent space.

Hamiltonian dynamics provides a second motivation. Ambrosio and Gangbo studied Hamiltonian evolutions in Wasserstein space \cite{Ambrosio+Gangbo_2008_Hamiltonian}; Chow, Li, and Zhou derived density--potential Hamiltonian and Euler--Lagrange equations for the classical Wasserstein metric \cite{Chow+Li+Zhou_2020_Wasserstein}. Newton equations on density spaces and diffeomorphism groups offer a related viewpoint \cite{Khesin+Misiolek+Modin_2021_Geometric}. Schr\"odinger problems connect this geometry with stochastic interpolation \cite{Leonard_2014_Survey}, Fisher-information acceleration, and Hamiltonian Hopf--Cole transformations \cite{Conforti_2019_Second,Gentil+Leonard+Ripani_2020_Dynamical,Leger+Li_2021_HopfCole}; graph transport also leads to discrete Schr\"odinger equations \cite{Chow+Li+Zhou_2019_Discrete}. For jump processes, Peletier, Rossi, Savar\'e, and Tse established a generalized gradient-flow framework containing a nonquadratic $\cosh$ dissipation with geometric-mean mobility \cite{Peletier+Rossi+Savare+Tse_2022_Jump}.
In the nonlocal continuum setting, however, the absence of a Levi-Civita calculus leaves a gap in the covariant
formulation of mechanical Hamiltonian flows and the 
coordinate-independent formulation of Euler--Lagrange equations.
The nonlocal Schr\"odinger problem offers a natural starting point,
but its Hopf--Cole transformation produces a strongly nonquadratic kinetic energy rather than a quadratic form directly
inducing the desired metric.

\paragraph{Main contributions.}
To address these gaps, we study nonlocal Wasserstein metrics on the space
$\mathcal P_+(M)$ of smooth strictly positive probability densities
on a closed Riemannian manifold $M$.
These metrics are induced by the Dirichlet forms associated with
the density-dependent jump kernels $J_\rho$ in
\eqref{eq:intro-weighted-kernel}.
Our main contributions are as follows.

\emph{The density-variation operator.}
To construct the Levi-Civita connection, we introduce the
density-variation operator $\Lambda_\rho$
(Definition~\ref{def:density-variation-operator}):
\[
\Lambda_\rho(\phi,\psi)(x)
:=
\int_M
\bigl(\phi(y)-\phi(x)\bigr)
\bigl(\psi(y)-\psi(x)\bigr)
\partial_a\theta(\rho(x),\rho(y))K(x,y)\,\dd y.
\]
This operator arises naturally in the density derivatives of the
Dirichlet form and the associated metric, and hence enters the
Levi-Civita Christoffel formula \eqref{eq:Christoffel-density}
through the Koszul formula. It thus takes over the role played by
the carr\'e du champ in the local Wasserstein calculus
\cite{Lott_2008_Geometric,Chow+Li+Zhou_2020_Wasserstein}.
Its diagonal value $\Lambda_\rho(\phi,\phi)$ also supplies
the kinetic term in the nonlocal Hamilton--Jacobi equation,
connecting covariant differentiation with Hamiltonian flows.

\emph{Two tangent representations and Levi-Civita covariant differentiation.}
We construct the density representation, based on density
variations with zero total mass, and the velocity-field
representation, based on nonlocal velocities
$\overline{\nabla}\phi(x,y):=\phi(y)-\phi(x)$.
We prove that the completed density tangent space and the closed
velocity-field tangent space are isometrically isomorphic.
We use the density-variation operator to derive explicit formulas for the
Levi-Civita connection and covariant differentiation in the density
representation, and then transfer to the velocity-field
representation via the isometry.
These formulas provide a nonlocal counterpart of the local
Wasserstein calculus
\cite{Lott_2008_Geometric}.
We also formulate gradient flows in both
representations, giving rise to a nonlocal continuity equation.
Examples for gradient flows include Boltzmann
and R\'enyi entropy flows, nonlocal mean-field equations,
and nonlocal Keller--Segel dynamics.

\emph{Hamiltonian--Newton structure.}
An action principle yields the Hamilton equations for a mechanical
Hamiltonian, coupling the nonlocal continuity equation with the
nonlocal Hamilton--Jacobi equation.
Using the Levi-Civita connection, we show that these equations are
equivalent to a nonlocal Newton equation with covariant acceleration.
These extend the local Wasserstein
framework of \cite{Chow+Li+Zhou_2020_Wasserstein} to nonlocal
metrics.
Given a smooth solution of a negative gradient flow under the
assumptions of
Proposition~\ref{prop:gradient-Hamiltonian-lift},
we construct for a gradient flow its zero-energy Hamiltonian representation, by lifting the potential energy to the metric norm of the driving
force's gradient.
Besides a nonlocal Vlasov equation
as an example for the Hamiltonian side, we also consider the nonlocal Schr\"odinger bridge
system, and the nonlocal Schr\"odinger equation.
For geometric-mean mobility $\theta(a,b)=\sqrt{ab}$,
the Hopf--Cole and Madelung transformations yield phase energies
involving $\cosh(\overline{\nabla}\phi)$ and
$\cos(\overline{\nabla}\phi)$, respectively.
Their second-order expansions at $\phi=0$ recover the same
kinetic energy that underlies our covariant calculus.
The hyperbolic phase energy shares the same $\cosh$ structure
of the dual dissipation potential studied for jump processes
in \cite{Peletier+Rossi+Savare+Tse_2022_Jump}.

\paragraph{Organization of the paper.}
Section~2 introduces the nonlocal operators, carr\'e du champ,
Dirichlet form, and potential spaces.
Section~3 presents the mobility-weighted calculus and density
variations of the Dirichlet form and kinetic energy.
Section~4 constructs the metrics in the two tangent representations,
establishes their equivalence, and derives gradients and energy dissipation.
Section~5 derives formulas for covariant differentiation and Hamiltonian dynamics,
including Legendre duality, the nonlocal Newton equation, and
Hamiltonian lifts of gradient flows.
Section~6 gives the gradient-flow and Hamiltonian examples.

\paragraph{Differentiation notations.}

Unless otherwise stated, the norm on a Banach space $X$ is denoted by
$\|\cdot\|_X$. For a Hilbert space $X$, we write
$\langle\cdot,\cdot\rangle_X$ for its inner product and
$\|u\|_X=\langle u,u\rangle_X^{1/2}$ for the induced norm.
Duality pairings are likewise distinguished by subscripts.

For a differentiable map $F:X\to Y$ between Banach spaces, we write
\begin{equation*}
\textstyle DF(u)[v]
:=\left.\frac{\mathrm d}{\mathrm d\epsilon}\right|_{\epsilon=0}
F(u+\epsilon v).
\end{equation*}
On Banach spaces, derivatives are understood in the Fr\'echet sense.

For a scalar functional, the variational derivative is characterized by
\begin{equation*}
\textstyle DF(u)[v]
=\left\langle\frac{\delta F}{\delta u}(u),v\right\rangle_{X^*,X}.
\end{equation*}
Here the subscript indicates Banach-space duality. For smooth function
variables, the $L^2(M)$ representation is
\begin{equation*}
\textstyle DF(u)[v]
=\left\langle\frac{\delta F}{\delta u}(u),v\right\rangle_{L^2(M)}
=\int_M\frac{\delta F}{\delta u}(u)(x)v(x)\,\mathrm dx.
\end{equation*}
On the probability-density manifold, variational derivatives are
defined up to additive spatial constants.

For a bivariate map $F:X\times Y\to Z$, we use
\[
\textstyle D_uF(u,w)[v]
:=\left.\frac{\mathrm d}{\mathrm d\epsilon}\right|_{\epsilon=0}
F(u+\epsilon v,w),
\qquad
D_wF(u,w)[z]
:=\left.\frac{\mathrm d}{\mathrm d\epsilon}\right|_{\epsilon=0}
F(u,w+\epsilon z).
\]
Thus $DF(u,w)[v,z]=D_uF(u,w)[v]+D_wF(u,w)[z]$.
For scalar-valued $F$, the corresponding variational derivatives satisfy
\[
\textstyle D_uF(u,w)[v]
=\left\langle\frac{\delta F}{\delta u}(u,w),v\right\rangle_{X^*,X},
\qquad
D_wF(u,w)[z]
=\left\langle\frac{\delta F}{\delta w}(u,w),z\right\rangle_{Y^*,Y}.
\]
Iterated derivatives are written as compositions or squares, such as
$D_uD_wF$ and $D_w^2F$, with their directions in brackets.
For density-indexed families, $DA_\rho[v]$ denotes the derivative
of $\rho\mapsto A_\rho$ at $\rho$ in direction $v$. Any operator
or bilinear-form arguments of $A_\rho$ are held fixed.

\section{Preliminaries}
\label{sec:preliminaries}

Let $M$ be a smooth connected closed
Riemannian manifold of positive finite dimension, with volume measure
$\mathrm dx$. Let $d_M$ denote the Riemannian distance on $M$. We use the same symbol for a measure and its density
with respect to the reference measure, whenever the latter exists.
For example, $\rho(\mathrm dx)=\rho(x)\,\mathrm dx$; on $M\times M$,
a nonnegative kernel $J$ also denotes the measure
\begin{equation*}
J(\mathrm dx,\mathrm dy):=J(x,y)\,\mathrm dx\,\mathrm dy.
\end{equation*}

All density, potential, and test fields used in the formal calculations
are $C^\infty$, also in time when time dependent. Jump kernels may be
measurable and singular. Energy expressions are evaluated on the
finite-energy domains introduced below, and operator expressions on
their smooth domains. 
Unless another measure is specified, $L^2(M)=L^2(M,\mathrm dx)$.

\subsection{Nonlocal operators}

The transitions between states are modeled by a jump kernel.

\begin{definition}[Jump kernel]
\label{def:symmetric-jump-kernel}
By a \emph{jump kernel}, we mean a measurable function $J:M\times M\to[0,\infty]$. In particular, it is called a \emph{symmetric jump kernel} if
\begin{equation*}
J(x,y)=J(y,x)
\end{equation*}
for $(\mathrm dx\otimes\mathrm dy)$-almost every $(x,y)\in M\times M$.
\end{definition}

Fractional-type generators have kernels proportional to
$|x-y|^{-d-2s}$, with $s\in(0,1)$.

To formulate fluid dynamics algebraically without relying on the chain rule, we define the nonlocal counterparts to classical differential operators.

\begin{definition}[Nonlocal differential operators]\label{def-nonlocal-differential-operators}
\quad
\begin{enumerate}
\item[(i)] For a scalar function
$\phi:M\to\mathbb R$, its nonlocal gradient (or graph gradient) is defined by
\begin{equation*}
\overline{\nabla}\phi(x,y):=\phi(y)-\phi(x).
\end{equation*}
		
\item[(ii)]
For a measurable edge field $F:M\times M\to\mathbb R\cup \{\pm\infty\}$, set
\[
F_\epsilon(x,y):=F(x,y)\mathbf 1_{\{d_M(x,y)>\epsilon\}},
\qquad \epsilon>0.
\]
Assume that $F_{\epsilon}\in L^1(M\times M)$ for every $\epsilon>0$ and define
\[
\overline{\operatorname{div}}F_\epsilon(x):=\int_{\{y\in M:\,d_M(x,y)>\epsilon\}}
\bigl(F(x,y)-F(y,x)\bigr)\,\mathrm dy.
\]
The nonlocal divergence (or graph divergence) of $F$ is defined by
\begin{equation}\label{eq:graph_div}
\overline{\operatorname{div}}F(x)
:=\lim_{\epsilon\downarrow0}\overline{\operatorname{div}}F_\epsilon(x)=\operatorname{p.v.}\int_M \bigl(F(x,y)-F(y,x)\bigr)\,\mathrm dy,
\end{equation}
provided that this limit exists for almost every $x$, belongs to $L^1(M)$, and satisfies
\[
\|\Div F_\epsilon-\Div F\|_{L^1(M)}\to0, \quad \epsilon\downarrow0.
\]
Here we use $\mathrm{p.v.}$ to mean that the distance-truncated integrals converge both almost everywhere and in $L^1(M)$ to the same limit.		
		
\item[(iii)] Let $J:M\times M\to[0,\infty]$ be a symmetric jump kernel. Set
\begin{equation*}
\begin{aligned}
\mathscr D(L_J):=\biggl\{\phi\in C^\infty(M):\,&
\iint_{M\times M}|\Nabla\phi|^2J\,\dd x\,\dd y<\infty, \, \forall \epsilon>0, (J\Nabla\phi)_\epsilon \in L^1(M\times M), \\
&\Div(J\Nabla\phi)\text{ is defined as in (ii)} \biggr\}.
\end{aligned}
\end{equation*}
For $\phi\in\mathscr D(L_J)$, define
\begin{equation*}
L_J\phi(x):=\Div(J\Nabla\phi)(x)
=\lim_{\epsilon\downarrow0}2\int_{\{y\in M:\,d_M(x,y)>\epsilon\}} \Nabla \phi(x,y) J(x,y) \dd y.
\end{equation*}
\end{enumerate}
\end{definition}

Note that, in particular,
$\overline{\nabla}\phi$ is an antisymmetric edge field.
For fields satisfying part (ii) of Definition~\ref{def-nonlocal-differential-operators}, the divergence depends only on the antisymmetric part of $F$. In particular, if $F$ is symmetric, i.e., $F(x,y) = F(y,x)$ for all $x,y \in M$, then $\Div F \equiv 0$.

The local version of $\Nabla$ is $\nabla$, of $\overline{\mathrm{div}}$ is $\mathrm{div}$, and of $L_J$ is $\Delta$.

At every distance cutoff the field is integrable on $M\times M$, so Fubini's theorem applies. The $L^1$ convergence in the definition then permits passage to the limit.

\begin{lemma}[Nonlocal integration-by-parts] \label{lemma:ibp}
Let $F$ satisfy Definition~\ref{def-nonlocal-differential-operators}-(ii),
and let $\phi\in C^\infty(M)$ satisfy $\iint_{M\times M}|\Nabla\phi(x,y)F(x,y)|\,\dd x\,\dd y<\infty$.
Then
\begin{equation}\label{eq:ibp}
\int_M\phi\,\Div F\,\dd x
=-\iint_{M\times M}\Nabla\phi\,F\,\dd x\,\dd y.
\end{equation}
\end{lemma}

\begin{proof}
Since $M$ is compact, $\phi$ is bounded. For each $\epsilon>0$, since
$F_\epsilon\in L^1(M\times M)$,
\[
\textstyle
\iint|\phi(x)F_\epsilon(x,y)|\,\dd x\,\dd y
\leq\|\phi\|_{L^\infty(M)}\|F_\epsilon\|_{L^1(M\times M)}<\infty.
\]
Fubini's theorem and exchange of $x$ and $y$ therefore give
\begin{align*}
\textstyle
\int_M\phi\,\Div F_\epsilon\,\dd x
&\textstyle=\iint\phi(x)
\bigl(F_\epsilon(x,y)-F_\epsilon(y,x)\bigr)\,\dd x\,\dd y\\
&\textstyle=-\iint\Nabla\phi(x,y)F_\epsilon(x,y)\,\dd x\,\dd y.
\end{align*}
On the left,
\[
\textstyle\left|\int_M\phi(\Div F_\epsilon-\Div F)\,\dd x\right|
\leq\|\phi\|_{L^\infty(M)}
\|\Div F_\epsilon-\Div F\|_{L^1(M)}\to0.
\]
On the right, $F_\epsilon\to F$ off the diagonal and
\(
|\Nabla\phi\,F_\epsilon|
\leq|\Nabla\phi\,F|\in L^1(M\times M).
\)
Dominated convergence now gives
\eqref{eq:ibp}.
\end{proof}

\begin{corollary}\label{cor:integral-LJ-zero}
For a symmetric jump kernel $J$ and $\phi\in\mathscr D(L_J)$,
\[
\textstyle\int_ML_J\phi\,\dd x=0.
\]
\end{corollary}

\begin{proof}
By the definition of $\mathscr D(L_J)$, the field $F=J\Nabla\phi$
satisfies Definition~\ref{def-nonlocal-differential-operators}-(ii).
Applying Lemma~\ref{lemma:ibp} to this field with test function $1$, the result follows.
\end{proof}


\subsection{Nonlocal carr\'e du champ}
The definition of the carr\'e du champ involves applying $L_J$ to a product. We therefore first establish the following property.

\begin{lemma}\label{lem:generator-domain-products}
If $\phi,\psi\in\mathscr D(L_J)$, then $\phi\psi\in\mathscr D(L_J)$.
\end{lemma}

\begin{proof}
Since $\phi,\psi$ are smooth and bounded,
\(
|\Nabla(\phi\psi)|
\leq\|\phi\|_{L^\infty(M)}|\Nabla\psi|
+\|\psi\|_{L^\infty(M)}|\Nabla\phi|.
\)
Thus 
\[
\textstyle\iint_{M\times M}|\Nabla(\phi\psi)|^2J\,\dd x\,\dd y<\infty,\quad (J\Nabla(\phi\psi))_\epsilon\in L^1(M\times M).
\]
The Cauchy--Schwarz inequality gives
\begin{equation}\label{eq:cross-energy-integrability}
\iint_{M\times M}|\Nabla\phi\,\Nabla\psi|J\,\dd x\,\dd y<\infty.
\end{equation}
For every $\epsilon>0$,
\begin{equation}\label{eq:distance-product-identity}
\begin{aligned}
\Div\bigl((J\Nabla(\phi\psi))_\epsilon\bigr)(x)
&=\phi(x)\Div\bigl((J\Nabla\psi)_\epsilon\bigr)(x)+\psi(x)\Div\bigl((J\Nabla\phi)_\epsilon\bigr)(x)\\
&\quad+2\int_{\{y:\,d_M(x,y)>\epsilon\}}
\Nabla\phi(x,y)\Nabla\psi(x,y)J(x,y)\,\dd y.
\end{aligned}
\end{equation}
The first two terms converge almost everywhere and in $L^1(M)$
by assumption. By \eqref{eq:cross-energy-integrability}, dominated
convergence gives the same convergence for the last term.
Hence $\phi\psi\in\mathscr D(L_J)$.
\end{proof}

The preceding product identity leads to the following definition, which measures the failure of $L_J$ to satisfy the Leibniz rule.

\begin{definition}[Nonlocal carr\'e du champ]
\label{def:nonlocal-carre-du-champ}
For a symmetric jump kernel $J$ and $\phi,\psi\in\mathscr D(L_J)$,
the carr\'e du champ associated with $L_J$ is defined by
\begin{equation*}
\Gamma_J(\phi,\psi):=\frac12\left[L_J(\phi\psi)-\phi L_J\psi-\psi L_J\phi\right].
\end{equation*}
\end{definition}

In this and the next subsection, let $J$ be a symmetric jump kernel
satisfying
\begin{equation}\label{eq:dirichlet-second-moment}
\iint_{M\times M}d_M(x,y)^2J(x,y)\,\dd x\,\dd y<\infty.
\end{equation}
Passing to the limit in \eqref{eq:distance-product-identity} gives
\begin{equation}
\label{eq:carre-du-champ-integral-representation}
\Gamma_J(\phi,\psi)(x)
=\int_M
\overline{\nabla}\phi(x,y)
\overline{\nabla}\psi(x,y)
J(x,y)\,\mathrm{d}y.
\end{equation}
By \eqref{eq:cross-energy-integrability} and Tonelli's theorem, \eqref{eq:carre-du-champ-integral-representation} holds and the Lebesgue integral in $y$ converges absolutely, for almost everywhere $x$. 
So no principal value is needed, and $\Gamma_J(\phi,\psi)\in L^1(M)$ for any
$\phi,\psi\in \mathscr D(L_J)$.

This integral expression also defines an $L^1(M)$ extension to any smooth finite-energy pair. 

Combining this integral representation with nonlocal integration by parts yields the following Green formula.



\begin{lemma}[Nonlocal Green's formula]
\label{lemma:nonlocal-Green}
Let $J$ be symmetric and $\phi,\psi\in\mathscr D(L_J)$. Then
\begin{equation}\label{nonlocal-Green}
\int_M\Gamma_J(\phi,\psi)\,\mathrm dx
=-\int_M\psi L_J\phi\,\mathrm dx
=-\int_M\phi L_J\psi\,\mathrm dx.
\end{equation}
\end{lemma}

\begin{proof}
Set $F=J\Nabla\phi$. By the definition of $\mathscr D(L_J)$,
$F_\epsilon\in L^1(M\times M)$ for every $\epsilon>0$,
and
\(
\Div F_\epsilon\to\Div F=L_J\phi
\)
in $L^1(M)$. And
\eqref{eq:cross-energy-integrability} verifies
\(
\iint_{M\times M}|\Nabla\psi\,F|\,\dd x\,\dd y
<\infty.
\)
Every hypothesis of Lemma~\ref{lemma:ibp} is therefore satisfied.
Applying that lemma gives
\[
\textstyle\int_M\psi L_J\phi\,\dd x
=-\iint_{M\times M}\Nabla\psi\,\Nabla\phi\,J\,\dd x\,\dd y.
\]
The integrand on the right is absolutely integrable by
\eqref{eq:cross-energy-integrability}. Fubini and
\eqref{eq:carre-du-champ-integral-representation} identify this double
integral with $\int_M\Gamma_J(\phi,\psi)\,\dd x$.
Interchanging $\phi$ and $\psi$ proves the other equality.
\end{proof}

From the proof we see that the first equality remains valid for $\phi\in\mathscr D(L_J)$ and any smooth finite-energy $\psi$. In this case $\Gamma_J(\phi,\psi)$ is understood by its integral expression, and no value of $L_J\psi$ is needed.

Restricting the domain to potentials $\phi\in\mathscr D(L_J)$ for which $L_J\phi\in L^2(M)$ gives a symmetric, non-positive operator in $L^2(M)$.

\subsection{Nonlocal Dirichlet form and potential space}
\label{subsubsec:nonlocal-Dirichlet-form-potential-space}

For $\phi,\psi\in C^\infty(M)$, 
\begin{equation}\label{eq:Dirichlet-form-definition}
\mathcal A_J(\phi,\psi)
:=\iint_{M\times M}\Nabla\phi(x,y)\Nabla\psi(x,y)
J(x,y)\,\dd x\,\dd y=\langle\Nabla\phi,\Nabla\psi\rangle_{L^2(M\times M,J)},
\end{equation}
Every smooth function is Lipschitz on the compact manifold, so
\eqref{eq:dirichlet-second-moment} makes \eqref{eq:Dirichlet-form-definition} finite.
For $\phi\in\mathscr D(L_J)$ and $\psi\in C^\infty(M)$,
the nonlocal Green formula \eqref{nonlocal-Green} gives
\begin{equation}\label{eq:Dirichlet-form-Green-L2}
\mathcal A_J(\phi,\psi)=-\int_M\psi L_J\phi\,\dd x.
\end{equation}

Thus $\mathcal A_J$ is a densely defined, symmetric, nonnegative
bilinear form on $L^2(M)$. When the nonlocal carr\'e du champ is defined,
\[
\mathcal A_J(\phi,\psi)=\int_M\Gamma_J(\phi,\psi)\,\dd x.
\]
Set $|\phi|_{\mathcal A_J}:=\mathcal A_J(\phi,\phi)^{1/2}$. We use the form norm and its completion
\begin{equation*}
\|\phi\|_{\mathscr D(\mathcal A_J)}^2
:=\|\phi\|_{L^2(M)}^2+|\phi|_{\mathcal A_J}^2,
\qquad
\mathscr D(\mathcal A_J)
:=\overline{C^\infty(M)}^{\,\|\cdot\|_{\mathscr D(\mathcal A_J)}}.
\end{equation*}
The completion is identified with a subspace of $L^2(M)$ and
\[
\mathscr D(L_J)\subseteq C^\infty(M)
\subseteq\mathscr D(\mathcal A_J).
\]

\begin{remark}[Dirichlet-form realization]
\label{lem:Dirichlet-form-realization}
This construction admits a rigorous realization as a regular symmetric Dirichlet
form on $L^2(M)$, with domain $\mathscr D(\mathcal A_J)$, core
$C^\infty(M)$, and the integral representation
\eqref{eq:Dirichlet-form-definition}
\cite{Fukushima+Oshima+Takeda_2011_Dirichlet}.
\end{remark}


The energy seminorm extends to $\mathscr D(\mathcal A_J)$, with
null space
\[
\mathscr N_J:=\{\phi\in\mathscr D(\mathcal A_J):
\mathcal A_J(\phi,\phi)=0\}.
\]
The Cauchy--Schwarz inequality
\begin{equation}\label{eq:energy-Cauchy-Schwarz}
|\mathcal A_J(\phi,\psi)|^2
\leq\mathcal A_J(\phi,\phi)\mathcal A_J(\psi,\psi)
\end{equation}
makes $\mathcal A_J$ an inner product on
$\mathscr D(\mathcal A_J)/\mathscr N_J$.

\begin{definition}[Homogeneous energy space of potentials]
\label{def:homogeneous-potential-energy-space}
The homogeneous energy space of potentials is the Hilbert completion
\[
\mathscr H_J
:=\overline{\mathscr D(\mathcal A_J)/\mathscr N_J}
^{\,|\cdot|_{\mathcal A_J}}.
\]
\end{definition}

Smooth functions are dense in the form norm in $\mathscr D(\mathcal A_J)$, so the
same energy completion is obtained from the smooth core modulo its
zero-energy elements. Unlike the form norm, this completion uses only
the energy seminorm.

The nonlocal gradient extends uniquely by completion to an isometric map
\begin{equation}\label{eq:isometric-gradient-map}
\Nabla:\mathscr H_J\to L^2(M\times M,J),
\qquad
\|\Nabla\phi\|_{L^2(M\times M,J)}=\|\phi\|_{\mathscr H_J}.
\end{equation}
The form also extends continuously to $\mathscr H_J$; for
representatives $\phi,\psi\in\mathscr D(\mathcal A_J)$,
\[
\langle[\phi],[\psi]\rangle_{\mathscr H_J}
=\mathcal A_J(\phi,\psi),\qquad
\|[\phi]\|_{\mathscr H_J}=|\phi|_{\mathcal A_J}.
\]

We call $J$ irreducible here if, for every
$\phi\in\mathscr D(\mathcal A_J)$,
\begin{equation}\label{eq:J-irreducibility}
\mathcal A_J(\phi,\phi)=0
\quad\Longrightarrow\quad
\phi\text{ is constant almost everywhere on }M.
\end{equation}
Then
\[
\mathscr N_J=\{c\mathbf1:c\in\mathbb R\}\simeq\mathbb R,
\qquad
\mathscr H_J
=\overline{\mathscr D(\mathcal A_J)/\mathbb R}
^{\,|\cdot|_{\mathcal A_J}}.
\]
Irreducibility makes representing potentials unique modulo constants
and the density metric nondegenerate on representable directions.
Solvability of the potential equation and smooth dependence on the
density require additional assumptions.

\paragraph{Convention.}
We suppress equivalence-class brackets, writing $\phi$ for $[\phi]$.
The Hilbert completion is retained for duality and continuous extensions;
its elements need not have smooth representatives. Pointwise and
differential calculations are performed on smooth representatives in
$\mathscr D(\mathcal A_J)$.

\subsection{Nonlocal kinetic energy}
\label{subsubsec:nonlocal-kinetic-energy}
The nonlocal kinetic energy $\mathcal{K}_J(\phi)$ takes a canonical quadratic form:
\begin{definition}[Nonlocal kinetic energy]
\label{def:nonlocal-kinetic-energy}
For every $\phi\in\mathscr H_J$, define
\begin{equation*}
\mathcal K_J(\phi)
:=\frac12\mathcal A_J(\phi,\phi)
=\frac12\|\phi\|_{\mathscr H_J}^2=\frac12\|\Nabla\phi\|_{L^2(M\times M,J)}^2.
\end{equation*}
\end{definition}

By the nonlocal Green's formula,
\begin{equation*}
\mathcal K_J(\phi)
=
-\frac12
\int_M
\phi(x)L_J\phi(x)\,\mathrm{d}x
\end{equation*}
whenever $\phi\in\mathscr D(L_J)$. In particular, $\mathcal K_J(\phi)\geq0$, since $L_J$ is non-positive.

\begin{lemma}[First and second variations of the kinetic energy]
\label{prop:kinetic-energy-variations}
Let $\phi,\psi,\xi\in\mathscr D(\mathcal A_J)$. Then
\begin{equation}
\label{eq:first-variation-kinetic-energy}
D\mathcal K_J(\phi)[\psi]=\mathcal A_J(\phi,\psi),
\end{equation}
and the second variation is
\begin{equation}
\label{eq:second-variation-kinetic-energy}
D^2\mathcal K_J(\phi)[\psi,\xi]=\mathcal A_J(\psi,\xi).
\end{equation}
\end{lemma}

\begin{proof}
Bilinearity gives the exact expansion
\begin{equation}
\label{eq:kinetic-energy-exact-expansion}
\mathcal K_J(\phi+\epsilon\psi)
=\mathcal K_J(\phi)+\epsilon\mathcal A_J(\phi,\psi)
+\frac{\epsilon^2}{2}\mathcal A_J(\psi,\psi),
\qquad \epsilon\in\mathbb R.
\end{equation}
Differentiating \eqref{eq:kinetic-energy-exact-expansion} at $\epsilon=0$
gives \eqref{eq:first-variation-kinetic-energy}.
Differentiating this expression in the $\phi$-direction $\xi$ yields
$\mathcal A_J(\xi,\psi)=\mathcal A_J(\psi,\xi)$, proving
\eqref{eq:second-variation-kinetic-energy}.
\end{proof}

\begin{remark}[Energy duality and $L^2$ representations]
\label{cor:kinetic-energy-variational-derivative}
\label{rem:weak-L2-variational-derivative}
Formula \eqref{eq:first-variation-kinetic-energy} extends continuously
to $\phi,\psi\in\mathscr H_J$.
By \eqref{eq:energy-Cauchy-Schwarz}, it defines
$D\mathcal K_J(\phi)=\mathcal A_J(\phi,\cdot)\in\mathscr H_J^*$,
even when $L_J\phi$ has no $L^2(M)$ representation.
If $\phi\in\mathscr D(L_J)$ and
$L_J\phi\in L^2(M)$, the nonlocal Green's formula gives
\begin{equation*}
D\mathcal K_J(\phi)[\psi]=-\langle L_J\phi,\psi\rangle_{L^2(M)},
\qquad \psi\in\mathscr D(\mathcal A_J).
\end{equation*}
Consequently, the $L^2(M)$ representation of this energy-dual derivative is
\begin{equation}
\label{eq:first-variation-kinetic-energy-Green}
\frac{\delta\mathcal K_J}{\delta\phi}(\phi)=-L_J\phi.
\end{equation}
\end{remark}

\begin{remark}[Convexity]
\label{rem:kinetic-energy-convexity}
Since
\(
D^2\mathcal K_J(\phi)[\psi,\psi]
=
\mathcal A_J(\psi,\psi)
\geq0,
\)
the functional $\mathcal K_J$ is convex. Its null directions are
precisely the elements of $\mathscr N_J$. Therefore,
$\mathcal K_J$ is strictly convex on
$\mathscr D(\mathcal A_J)/\mathscr N_J$. If $J$ is irreducible, then
$\mathscr N_J=\mathbb R$, and the kinetic energy is strictly convex
modulo additive constants.
\end{remark}

\section{Nonlocal calculus on Wasserstein space}

We denote by
\[
\mathcal P_+(M)
:=
\left\{
\rho\in C^\infty(M):
\rho>0\ \text{on }M,
\int_M\rho(x)\,\mathrm dx=1
\right\}
\]
the space of strictly positive probability densities with respect to
$\mathrm dx$. We regard $\mathcal P_+(M)$ as the formal density manifold
underlying Wasserstein space; nonlocality enters through the metric on its
tangent spaces.

\subsection{Mobility-weighted jump kernels}
\label{subsubsec:mobility-weighted-jump-kernels}

To construct a nonlocal metric on Wasserstein space, we use a density-dependent
jump kernel. Its dependence on $\rho$ is specified by a mobility function.

\begin{definition}[Mobility]
Let
\(
\theta:(0,\infty)\times(0,\infty)\to(0,\infty)
\)
be a $C^\infty$-map that is symmetric, i.e.,
\(
\theta(a,b)=\theta(b,a),
 a,b>0.
\)
For every strictly positive density $\rho$, a function of the form
\begin{equation*}
\theta_\rho(x,y)
:=
\theta\bigl(\rho(x),\rho(y)\bigr),
\end{equation*}
is called a \emph{mobility}.
\end{definition}

By the symmetry of $\theta$,
\begin{equation}\label{eq:partial-symmetry}
\partial_a\theta(a,b)
=
\partial_b\theta(b,a).
\end{equation}

For $\sigma:M\to\mathbb R$ with $\int_M \sigma(x)\,\dd x = 0$, the directional derivative of the mobility is
\begin{equation}\label{eq:general-mobility-variation}
D\theta_\rho[\sigma](x,y)
=
\partial_a\theta\bigl(\rho(x),\rho(y)\bigr)\sigma(x) +
\partial_b\theta\bigl(\rho(x),\rho(y)\bigr)\sigma(y).
\end{equation}
By \eqref{eq:partial-symmetry}, $D\theta_\rho[\sigma]$ is also symmetric:
\begin{equation*}
D\theta_\rho[\sigma](x,y)
=
D\theta_\rho[\sigma](y,x).
\end{equation*}

\begin{example}[Geometric mean and logarithmic mean]
Two distinguished examples of the symmetric function $\theta$ are provided
by the geometric mean and logarithmic mean: for $a,b>0$,
\begin{align}
\text{geometric mean:} \quad & \theta_{\mathrm{geom}}(a,b):=\sqrt{ab}, \label{eq:geometric-mean-definition} \\
\text{logarithmic mean:} \quad & \theta_{\log}(a,b)
:=
\begin{cases}
\frac{b-a}{\log b-\log a},
& a\neq b,
\\
a,
& a=b.
\end{cases} \label{eq:logarithmic-mean-definition}
\end{align}
Both means belong to $C^\infty((0,\infty)^2)$.
For later use, we compute the derivatives of the geometric mean:
\begin{equation*}
\textstyle\partial_a\theta_{\mathrm{geom}}(a,b)=\frac{1}{2}\sqrt{\frac ba},\quad\partial_b\theta_{\mathrm{geom}}(a,b)=\frac{1}{2}\sqrt{\frac ab}.
\end{equation*}
Consequently,
\begin{equation*}
\textstyle D(\theta_{\mathrm{geom}})_\rho[\sigma](x,y)=\frac{1}{2}\sqrt{\rho(x)\rho(y)}\left(\frac{\sigma(x)}{\rho(x)}+\frac{\sigma(y)}{\rho(y)}\right).
\end{equation*}
\end{example}

Let
$
\theta
$
be a mobility.
Let
\(
K:M\times M\rightarrow[0,\infty]
\)
be a symmetric kernel, namely, referred to as the reference kernel.

\begin{assumption}[Reference kernel]
\label{ass:reference-kernel}
Throughout the remainder of the paper, assume that the symmetric reference kernel
$K$ is irreducible (in the sense of \eqref{eq:J-irreducibility}) and satisfies
\begin{equation*}
\iint_{M\times M}d_M(x,y)^2K(x,y)\,\dd x\,\dd y<\infty.
\end{equation*}
\end{assumption}
For every strictly positive density
$\rho\in\mathcal P_+(M)$, define the corresponding mobility-weighted jump kernel by
\begin{equation*}
J_\rho(x,y):=\theta_\rho(x,y)K(x,y).
\end{equation*}
By the symmetry of $K$ and $\theta$, one has
\(
J_\rho(x,y)=J_\rho(y,x)
\)
for almost every $(x,y)\in M\times M$. Consequently, for each fixed
density $\rho$, all the constructions introduced above for a general
symmetric jump kernel apply to $J_\rho$ on their stated domains.
In operator formulas,
$\mathscr D(L_\rho)$ denotes the domain specified in
Definition~\ref{def-nonlocal-differential-operators} with $J=J_\rho$.
The realization on $\mathscr H_\rho$ is defined through the
Dirichlet-form pairing.

In the sequel, we use the abbreviations
\[
L_\rho:=L_{J_\rho},\quad\Gamma_\rho:=\Gamma_{J_\rho},
\quad\mathcal A_\rho:=\mathcal A_{J_\rho}.
\]
Here the density $\rho$ is held fixed.

\begin{remark}[Equivalence of energy spaces]
Because $\rho(M)$ is a compact subset of $(0,\infty)$ and $\theta$ is smooth and positive,
there are $0<c_\rho\leq C_\rho<\infty$ such that
\[
c_\rho\leq\theta_\rho(x,y)\leq C_\rho,
\qquad
c_\rho\mathcal A_K
\leq\mathcal A_\rho
\leq C_\rho\mathcal A_K
\]
Consequently, the forms $\mathcal A_\rho$ and $\mathcal A_K$ have the same domain and null space,
and their homogeneous energy completions are canonically identified.
In particular, irreducibility of $K$ in Assumption~\ref{ass:reference-kernel} passes to $J_\rho$.
\end{remark}

\paragraph{Potential spaces associated with $J_\rho$.}

The symmetric edge measure associated with $J_\rho$ is
\begin{equation*}
J_\rho(\mathrm dx,\mathrm dy)
=J_\rho(x,y)\,\mathrm dx\,\mathrm dy.
\end{equation*}
We write
\begin{equation*}
\mathscr D(\mathcal A_\rho):=\mathscr D(\mathcal A_{J_\rho}),
\quad
\mathscr N_\rho:=\mathscr N_{J_\rho},
\quad
\mathscr H_\rho:=\mathscr H_{J_\rho}.
\end{equation*}
Every $\phi\in\mathscr D(\mathcal A_\rho)$ has
$\Nabla\phi\in L^2(M\times M,J_\rho)$ by the closed-form
integral representation. In particular,
$C^\infty(M)\subseteq\mathscr D(\mathcal A_\rho)$.
Its energy seminorm is
\(
|\phi|_{\mathcal A_\rho}^2
:=\mathcal A_\rho(\phi,\phi).
\)
The homogeneous potential space is
\begin{equation*}
\mathscr H_\rho
=\overline{\mathscr D(\mathcal A_\rho)/\mathscr N_\rho}^{\,|\cdot|_{\mathcal A_\rho}}.
\end{equation*}



\subsection{Density variation of the Dirichlet form}
\label{subsubsec:density-variation-of-the-dirichlet-form}

We first differentiate the Dirichlet form with respect to the density,
then obtain the kinetic-energy variations as a corollary.

\begin{definition}[Density-variation operator]
\label{def:density-variation-operator}
For $\rho\in\mathcal P_+(M)$ and
$\phi,\psi\in\mathscr D(\mathcal A_\rho)$, define the
\emph{density-variation operator} $\Lambda_\rho$ almost everywhere by
\begin{equation}\label{eq:density-variation-operator}
\Lambda_\rho(\phi,\psi)(x)
:=\int_M\Nabla\phi(x,y)\Nabla\psi(x,y)
\partial_a\theta(\rho(x),\rho(y))K(x,y)\,\dd y.
\end{equation}
\end{definition}
Since $\rho(M)$ is a compact subset of $(0,\infty)$ and
$\theta$ is smooth and positive, there exists $C_\rho>0$ such that $\frac{|\partial_a\theta(\rho(x),\rho(y))|}{\theta_\rho(x,y)}=|\partial_a\log\theta(\rho(x),\rho(y))|
\leq C_\rho$. For $\phi,\psi\in\mathscr D(\mathcal A_\rho)$,
the Cauchy--Schwarz inequality therefore gives
\[
\|\Lambda_\rho(\phi,\psi)\|_{L^1(M)}
\leq C_\rho\iint|\Nabla\phi\,\Nabla\psi|J_\rho\,\dd x\,\dd y
\leq C_\rho\|\phi\|_{\mathscr H_\rho}\|\psi\|_{\mathscr H_\rho}.
\]
Since $|\partial_a\theta(\rho(x),\rho(y))|\le C_\rho\theta(\rho(x),\rho(y))$, the Cauchy--Schwarz inequality shows that the integrand in \eqref{eq:density-variation-operator} is absolutely integrable on $M\times M$. Tonelli's theorem then implies that the Lebesgue integral converges absolutely for almost every $x$.
Thus $\Lambda_\rho$ extends continuously to
$\mathscr H_\rho\times\mathscr H_\rho$ with values in $L^1(M)$.
It describes density variation of the form and is distinct from
the carr\'e du champ $\Gamma_\rho$.

\begin{lemma}[Density variation of the Dirichlet form]
\label{lem:density-variation-dirichlet-form}
For a fixed smooth zero-mass $\sigma$ and $\phi,\psi\in\mathscr H_\rho$,
the density-variation operator satisfies
\begin{equation}\label{eq:key-duality}
\begin{aligned}
D\mathcal A_\rho[\sigma](\phi,\psi)
&=\iint_{M\times M}\Nabla\phi\,\Nabla\psi\,
D\theta_\rho[\sigma]K\,\dd x\,\dd y\\
&=2\int_M\sigma\,\Lambda_\rho(\phi,\psi)\,\dd x,
\end{aligned}
\end{equation}
and $D\mathcal A_\rho[\sigma]$ is a symmetric bilinear form on $\mathscr H_\rho$.
\end{lemma}

\begin{proof}
For small $\epsilon$, $\rho+\epsilon\sigma$ takes values in a fixed compact
subset of $(0,\infty)$. The mean value theorem gives
\[
\textstyle\left|\frac{\theta_{\rho+\epsilon\sigma}-\theta_\rho}{\epsilon}\right|
\leq C\theta_\rho,
\qquad
|\partial_i\theta(\rho(x),\rho(y))|\leq C\theta_\rho(x,y),
\quad i=1,2.
\]
Since
\[
\textstyle
\iint|\Nabla\phi\,\Nabla\psi|\theta_\rho K\,\dd x\,\dd y
\leq\|\phi\|_{\mathscr H_\rho}\|\psi\|_{\mathscr H_\rho},
\]
dominated convergence permits differentiation under the integral.
Using \eqref{eq:general-mobility-variation},
\[
\begin{aligned}
D\mathcal A_\rho[\sigma](\phi,\psi)
&=\iint\Nabla\phi\,\Nabla\psi\,
\partial_a\theta(\rho(x),\rho(y))\sigma(x)K\,\dd x\,\dd y\\
&\quad+\iint\Nabla\phi\,\Nabla\psi\,
\partial_b\theta(\rho(x),\rho(y))\sigma(y)K\,\dd x\,\dd y\\
&=2\iint\Nabla\phi\,\Nabla\psi\,
\partial_a\theta(\rho(x),\rho(y))\sigma(x)K\,\dd x\,\dd y\\
&=2\int\sigma\,\Lambda_\rho(\phi,\psi)\,\dd x.
\end{aligned}
\]
Both double integrals are absolutely convergent; exchanging $x,y$
in the second and using \eqref{eq:partial-symmetry} makes them equal. The last equality follows from \eqref{eq:density-variation-operator} and Fubini’s theorem.
\end{proof}

We write $\langle\psi,\sigma\rangle_\rho$ for the potential--density
duality pairing. More generally,
$\langle\psi,\ell\rangle_\rho:=\ell(\psi)$ for
$\ell\in\mathscr H_\rho^*$, the continuous dual of $\mathscr H_\rho$.
The operator $-L_\rho:\mathscr H_\rho\to\mathscr H_\rho^*$ is defined
by the Dirichlet-form pairing
\begin{equation}
\label{eq:weak-operator-dual-definition}
\langle\psi,-L_\rho\phi\rangle_\rho
:=\mathcal A_\rho(\phi,\psi),
\qquad \phi,\psi\in\mathscr H_\rho.
\end{equation}
Continuity follows from \eqref{eq:energy-Cauchy-Schwarz}.

\begin{remark}[Operator representation of the density variation]
\label{rem:generator-density-variation}
With the realization \eqref{eq:weak-operator-dual-definition},
the density derivative of $L_\rho$ satisfies
\begin{equation*}
\textstyle-\langle\psi,(DL_\rho[\sigma])\phi\rangle_\rho
=D\mathcal A_\rho[\sigma](\phi,\psi)
=2\int_M\sigma\,\Lambda_\rho(\phi,\psi)\,\dd x.
\end{equation*}
If $\phi\in\mathscr D(L_{\rho})$, then
$\phi\in\mathscr D(L_{\rho+\epsilon\sigma})$ for small $\epsilon$, and
$\epsilon\mapsto L_{\rho+\epsilon\sigma}\phi$ is differentiable in $L^1(M)$, with
\begin{equation*}
(DL_\rho[\sigma])\phi(x)
=\Div\bigl(D\theta_\rho[\sigma]K\Nabla\phi\bigr)(x),
\end{equation*}
\end{remark}

For $\rho\in\mathcal P_+(M)$ and $\phi\in\mathscr H_\rho$,
Definition~\ref{def:nonlocal-kinetic-energy} applied to $J_\rho$ and we denote
\[
\mathcal K(\rho,\phi)
:=\mathcal K_{J_\rho}(\phi)
=\tfrac12\mathcal A_\rho(\phi,\phi)=\tfrac12\|\Nabla\phi\|_{L^2(M\times M,J_\rho)}^2.
\]

At fixed $\rho$, Lemma~\ref{prop:kinetic-energy-variations} with
$J=J_\rho$ gives the potential variation. At fixed $\phi$,
differentiating $\mathcal K=\tfrac12\mathcal A_\rho(\phi,\phi)$
and applying Lemma~\ref{lem:density-variation-dirichlet-form} with
$\psi=\phi$ gives the density variation. Thus:

\begin{corollary}[Variations of the nonlocal kinetic energy]
\label{cor:kinetic-density-variation}
For $\phi,\psi\in\mathscr H_\rho$ and smooth zero-mass $\sigma$,
\begin{equation*}
\begin{aligned}
D_\phi\mathcal K(\rho,\phi)[\psi]&=\mathcal A_\rho(\phi,\psi),\\
D_\rho\mathcal K(\rho,\phi)[\sigma]
&=\int_M\sigma\,\Lambda_\rho(\phi,\phi)\,\dd x.
\end{aligned}
\end{equation*}
\end{corollary}

\begin{remark}[Operator representation of the kinetic energy]
\label{rem:nonlocal-kinetic-energy-operator}
These directional derivatives identify the corresponding variational
derivatives. The potential derivative belongs to $\mathscr H_\rho^*$,
whereas the density derivative is represented by an $L^1(M)$ function
determined modulo constants, since density variations have zero mass.
Using \eqref{eq:weak-operator-dual-definition}, we obtain
\begin{equation*}
\begin{aligned}
\frac{\delta\mathcal K}{\delta\phi}(\rho,\phi)
&=\mathcal A_\rho(\phi,\cdot)=-L_\rho\phi\in\mathscr H_\rho^*,\\
\frac{\delta\mathcal K}{\delta\rho}(\rho,\phi)
&=\Lambda_\rho(\phi,\phi)\qquad\text{modulo spatial constants}.
\end{aligned}
\end{equation*}

For $\phi\in\mathscr D(L_\rho)$, nonlocal Green's formula gives
\[
\mathcal K(\rho,\phi)=-\frac12\int_M\phi L_\rho\phi\,\dd x,
\qquad
D_\phi\mathcal K(\rho,\phi)[\psi]
=-\int_M\psi L_\rho\phi\,\dd x
\]
for smooth $\psi$. If also $L_\rho\phi\in L^2(M)$,
\eqref{eq:first-variation-kinetic-energy-Green} with $J=J_\rho$
identifies $-L_\rho\phi$ as the $L^2(M)$ representative of the
variational derivative; the latter pairing identity extends to
$\psi\in\mathscr D(\mathcal A_\rho)$ by
\eqref{eq:Dirichlet-form-Green-L2}.
\end{remark}

\section{Nonlocal metric and gradient flows}
\label{sec:riemannian-structure-gradient-flows}

We construct a nonlocal metric on Wasserstein space using
the potential spaces and quotient convention of
Section~\ref{subsubsec:mobility-weighted-jump-kernels}. 
The construction applies to any symmetric positive mobility $\theta$.
We first construct the density tangent structure and its metric, then the
velocity-field representation. Their isometric identification leads to
equivalent descriptions of gradients and gradient flows for the nonlocal metric.

\subsection{Density tangent structure and nonlocal metric}
\label{subsec:general-mobility}

\begin{definition}[Density-coordinate tangent space]
\label{def:density-coordinate-tangent-space}
The tangent space in density coordinates is
\begin{equation*}
T_\rho^{\mathrm{dens}}\mathcal P_+(M)
:=\left\{\sigma\in C^\infty(M):\int_M\sigma\,\mathrm dx=0\right\}.
\end{equation*}
\end{definition}

\begin{assumption}[Smooth realization]
\label{ass:smooth-realization}
In addition to Assumption~\ref{ass:reference-kernel}, throughout the
remainder of the paper we assume:
\begin{enumerate}
\item[(1)] For every $\rho\in\mathcal P_+(M)$ and every $\phi\in\mathscr D(L_\rho)$,
$L_\rho\phi\in C^\infty(M)$;
the map $(\rho,\phi)\mapsto L_\rho\phi$ is jointly smooth.
\item[(2)] For every $\rho\in\mathcal P_+(M)$ and every smooth
$\sigma$ with $\int_M\sigma\,\dd x=0$, the normalized pseudoinverse gives
\[
\phi=(-L_\rho)^\dagger\sigma\in C^\infty(M),\qquad
-L_\rho\phi=\sigma,\qquad \int_M\phi\,\dd x=0.
\]
The map $(\rho,\sigma)\mapsto(-L_\rho)^\dagger\sigma$ is jointly smooth.
\item[(3)] The density-variation operator in \eqref{eq:density-variation-operator} is jointly smooth
as a map $(\rho,\phi,\psi)\mapsto\Lambda_\rho(\phi,\psi)$,
with values in $C^\infty(M)$ for smooth potentials $\phi,\psi$.
\end{enumerate}
Smoothness is understood in the usual $C^\infty$ topologies: all
iterated directional derivatives exist and are jointly continuous.
\end{assumption}

\begin{remark}[Pseudoinverse and normalized inverse]
\label{rem:normalized-inverse}
Under Assumptions~\ref{ass:reference-kernel} and~\ref{ass:smooth-realization}, $\ker L_\rho=\mathbb R$.
Thus $(-L_\rho)^\dagger$ is the genuine inverse of $-L_\rho$ between
smooth zero-mean potentials and smooth zero-mass density vectors; on
arbitrary smooth potentials,
\[
\textstyle(-L_\rho)^\dagger(-L_\rho)\phi
=\phi-\frac{1}{\operatorname{vol}(M)}\int_M\phi\,\dd x.
\]
\end{remark}

We also write $T_\rho\mathcal P_+(M)$ when the representation is clear.
For a density tangent vector $\sigma$, let $\phi$ solve
\begin{equation*}
\sigma=-L_\rho\phi.
\end{equation*}
We call $\phi$ the \emph{potential representing $\sigma$}, or its
\emph{velocity potential}, since it generates the nonlocal velocity
$\Nabla\phi$.
Irreducibility from Assumption~\ref{ass:reference-kernel} determines $\phi$ modulo spatial constants.
We choose representatives with a fixed normalization and write
\begin{equation}
\label{eq:general-potential-pseudoinverse}
\phi=-L_\rho^\dagger\sigma=(-L_\rho)^\dagger\sigma,
\end{equation}
where $L_\rho^\dagger$ denotes the pseudoinverse of $L_\rho$ modulo constants.

We use $\sigma,\tau,\eta$ for density tangent vectors and
$\phi,\psi,\xi$ for their respective potentials.

\begin{definition}[Density-coordinate metric]
\label{def:density-coordinate-metric}
For $\sigma,\tau\in T_\rho^{\mathrm{dens}}\mathcal P_+(M)$ with
respective potentials $\phi,\psi$ satisfying
$\sigma=-L_\rho\phi$ and $\tau=-L_\rho\psi$, define
\begin{equation*}
G_\rho(\sigma,\tau)
:=\mathcal A_\rho(\phi,\psi),
\qquad
\|\sigma\|_{G_\rho}:=G_\rho(\sigma,\sigma)^{1/2}.
\end{equation*}
\end{definition}
This is independent of the representatives of the potentials.

\begin{lemma}[Equivalent representations of the density metric]
\label{prop:density-metric-representations}
Let $\sigma,\tau$ be smooth density tangent vectors with
representing potentials $\phi,\psi\in\mathscr D(L_\rho)$,
$\sigma=-L_\rho\phi$ and $\tau=-L_\rho\psi$. The metric satisfies
\begin{equation}
\label{eq:density-duality}
G_\rho(\sigma,\tau)
=\int_M\sigma\psi\,\mathrm dx
=\int_M\phi\tau\,\mathrm dx=\int_M\sigma(-L_\rho)^\dagger\tau\,\mathrm dx.
\end{equation}
In particular,
\begin{equation*}
G_\rho(\sigma,\sigma)
=\|\Nabla\phi\|_{L^2(M\times M,J_\rho)}^2,
\end{equation*}
and $G_\rho(\sigma,\sigma)=0$ if and only if $\sigma=0$.
\end{lemma}

\begin{proof}
By the nonlocal Green's formula and
$\tau=-L_\rho\psi$,
\[
G_\rho(\sigma,\tau)
=\mathcal A_\rho(\phi,\psi)
=-\int_M\phi L_\rho\psi\,\mathrm dx
=\int_M\phi\tau\,\mathrm dx.
\]
Similarly, using $\sigma=-L_\rho\phi$ gives
$G_\rho(\sigma,\tau)=\int_M\sigma\psi\,\mathrm dx$.
Substituting $\psi=-L_\rho^\dagger\tau=(-L_\rho)^\dagger\tau$
yields the remaining expressions in \eqref{eq:density-duality}.
Taking $\tau=\sigma$ gives
\[
G_\rho(\sigma,\sigma)
=\mathcal A_\rho(\phi,\phi)
=\|\Nabla\phi\|_{L^2(M\times M,J_\rho)}^2.
\]
If this quantity vanishes, irreducibility makes $\phi$
constant, hence $\sigma=-L_\rho\phi=0$; the converse is immediate.
\end{proof}

\subsubsection*{Cotangent representation and musical maps}

Let $T_\rho^*\mathcal P_+(M)$ denote the continuous dual of the
density tangent space equipped with $\|\cdot\|_{G_\rho}$.

Recall the potential--density pairing and the Dirichlet-form
realization \eqref{eq:weak-operator-dual-definition}.
For $\phi\in\mathscr D(L_\rho)$, $\sigma=-L_\rho\phi$,
and a smooth finite-energy $\psi$, the one-sided Green formula gives
\begin{equation*}
\langle\psi,\sigma\rangle_\rho
=\mathcal A_\rho(\phi,\psi)
=\langle\psi,\sigma\rangle_{L^2(M)},
\qquad \sigma=-L_\rho\phi.
\end{equation*}

\begin{lemma}[Potential representation of the cotangent space]
\label{prop:density-cotangent-space}
Every $\ell\in T_\rho^*\mathcal P_+(M)$ is represented by a unique
$\psi\in\mathscr H_\rho$ through
\[
\ell(\sigma)=\langle\psi,\sigma\rangle_\rho
=\mathcal A_\rho(\phi,\psi),
\qquad \sigma=-L_\rho\phi.
\]
This correspondence gives the isometric identification
\begin{equation}
\label{eq:potential-cotangent-space}
T_\rho^*\mathcal P_+(M)\simeq\mathscr H_\rho,
\qquad
\|\ell\|_{T_\rho^*\mathcal P_+(M)}=\|\psi\|_{\mathscr H_\rho}.
\end{equation}
\end{lemma}

\begin{proof}
For $\psi\in\mathscr H_\rho$, set
$\ell_\psi(\sigma):=\mathcal A_\rho(\phi,\psi)$, where
$\sigma=-L_\rho\phi$. By the Cauchy--Schwarz inequality \eqref{eq:energy-Cauchy-Schwarz},
\[
|\ell_\psi(\sigma)|
\leq\|\phi\|_{\mathscr H_\rho}\|\psi\|_{\mathscr H_\rho}
=\|\sigma\|_{G_\rho}\|\psi\|_{\mathscr H_\rho},
\]
so $\ell_\psi$ is continuous.
Conversely, given $\ell\in T_\rho^*\mathcal P_+(M)$, define
$F(\phi):=\ell(-L_\rho\phi)$ on the potentials representing smooth density tangent vectors. Then
\[
|F(\phi)|
\leq\|\ell\|_{T_\rho^*\mathcal P_+(M)}\|L_\rho\phi\|_{G_\rho}
=\|\ell\|_{T_\rho^*\mathcal P_+(M)}\|\phi\|_{\mathscr H_\rho}.
\]
By denseness, $F$ extends uniquely to a continuous functional on
$\mathscr H_\rho$. The Riesz representation theorem yields a unique
$\psi\in\mathscr H_\rho$ such that
\[
\ell(\sigma)=F(\phi)
=\mathcal A_\rho(\phi,\psi)=\langle\psi,\sigma\rangle_\rho,
\qquad \sigma=-L_\rho\phi.
\]
Finally, taking the supremum over unit-energy representing potentials gives
\[
\|\ell_\psi\|_{T_\rho^*\mathcal P_+(M)}
=\sup_{\substack{\phi\in C^{\infty}(M),\
-L_\rho\phi\in T_\rho^{\mathrm{dens}}\mathcal P_+(M)\\
\|\phi\|_{\mathscr H_\rho}\leq1}}
|\mathcal A_\rho(\phi,\psi)|
=\|\psi\|_{\mathscr H_\rho},
\]
again by denseness.
\end{proof}

In the musical-map statements immediately below,
$T_\rho^{\mathrm{dens}}\mathcal P_+(M)$ denotes the $G_\rho$-completion
of the smooth density tangent space. Its continuous dual is unchanged.
By denseness, $-L_\rho$, defined through energy duality, extends to an isometric
isomorphism from $\mathscr H_\rho$ onto this completion. Its inverse is the continuous extension of $(-L_\rho)^\dagger$ to this completion. All pairings
in these statements are energy dualities, and
\eqref{eq:potential-cotangent-space} remains valid. We return to
smooth tangent vectors for the subsequent differentiation arguments.

For fixed $\rho$, the \emph{metric musical maps} are the flat map
$\flat:T_\rho^{\mathrm{dens}}\mathcal P_+(M)\to T_\rho^*\mathcal P_+(M)$,
characterized by
\[
\langle\sigma^\flat,\tau\rangle_\rho=G_\rho(\sigma,\tau),
\]
and its inverse, the \emph{sharp map} $\sharp=\flat^{-1}$.

\begin{lemma}[Musical maps in density coordinates]
\label{cor:metric-musical-maps}
For $\sigma=-L_\rho\phi$, the flat and sharp maps are
\begin{equation}
\label{eq:metric-Legendre-maps}
\sigma^\flat=(-L_\rho)^\dagger\sigma=\phi,
\qquad
\phi^\sharp=-L_\rho\phi=\sigma.
\end{equation}
\end{lemma}

\begin{proof}
By continuous extension of the smooth identity in
Lemma~\ref{prop:density-metric-representations}, for every completed
density tangent vector $\tau$,
\[
\langle\sigma^\flat,\tau\rangle_\rho
=G_\rho(\sigma,\tau)
=\langle\phi,\tau\rangle_\rho.
\]
Thus $\sigma^\flat=\phi$ as a cotangent vector, and taking the inverse
gives $\phi^\sharp=-L_\rho\phi$.
\end{proof}

\subsubsection*{Variation of the representing potential and the metric}

Here and in the following metric-variation arguments, density
vectors and their normalized representing potentials are smooth.
Assumption~\ref{ass:smooth-realization} makes the inverse map
smooth in the $C^\infty$ topology. Its derivatives are therefore
smooth potentials in $\mathscr D(L_\rho)$, and the product rule below
is justified in $C^\infty(M)$.

When $\eta$ is held fixed in density coordinates, its representing
potential $\xi$ still depends on $\rho$. In $(DL_\rho[\sigma])\xi$,
only the operator is differentiated and $\xi$ is frozen at its
current value. Differentiating the product $L_\rho\xi(\rho)$ also
includes the variation of $\xi$.

For fixed $\eta$, the potential equation along $\rho+\epsilon\sigma$ is
\[
\eta=-L_{\rho+\epsilon\sigma}\xi(\rho+\epsilon\sigma).
\]
Differentiating at $\epsilon=0$ and using the product rule, we obtain
\[
\begin{aligned}
0
&=-\left.\frac{\mathrm d}{\mathrm d\epsilon}\right|_{\epsilon=0}
\bigl(L_{\rho+\epsilon\sigma}\xi(\rho+\epsilon\sigma)\bigr)\\
&=-(DL_\rho[\sigma])\xi-L_\rho D\xi(\rho)[\sigma].
\end{aligned}
\]
Rearranging gives
\begin{equation}
\label{eq:potential-variation-fixed-vector}
L_\rho D\xi(\rho)[\sigma]=-(DL_\rho[\sigma])\xi.
\end{equation}
Applying the pseudoinverse therefore gives, with the chosen normalization,
\begin{equation*}
D\xi(\rho)[\sigma]
=(-L_\rho)^\dagger(DL_\rho[\sigma])\xi.
\end{equation*}


\begin{lemma}[Variation of the density-coordinate metric]
\label{lem:density-metric-variation}
For fixed density tangent vectors $\eta=-L_\rho\xi$ and
$\tau=-L_\rho\psi$,
\begin{equation}
\label{eq:density-metric-variation}
DG_\rho[\sigma](\eta,\tau)
=-2\int_M\sigma\,\Lambda_\rho(\xi,\psi)\,\mathrm dx.
\end{equation}
\end{lemma}

\begin{proof}
Keep $\eta,\tau$ fixed. Differentiating the duality formula
$G_\rho(\eta,\tau)=\int_M\eta\psi\,\mathrm dx$ gives
\begin{align*}
DG_\rho[\sigma](\eta,\tau)
&=\int_M\eta D\psi(\rho)[\sigma]\,\mathrm dx\\
&=-\int_M(L_\rho\xi)D\psi(\rho)[\sigma]\,\mathrm dx\\
&=-\int_M\xi L_\rho(D\psi(\rho)[\sigma])\,\mathrm dx\\
&=\int_M\xi(DL_\rho[\sigma])\psi\,\mathrm dx\\
&=-2\int_M\sigma\,\Lambda_\rho(\xi,\psi)\,\mathrm dx.
\end{align*}
The second equality uses $\eta=-L_\rho\xi$; the third
uses symmetry of $L_\rho$. The fourth follows from \eqref{eq:potential-variation-fixed-vector};
the last uses \eqref{eq:key-duality} and the symmetry of
$\Lambda_\rho$.
\end{proof}

This variation fixes the density tangent vectors, so their representing
potentials vary with $\rho$. The fixed-potential variation will be
considered in Section~\ref{subsubsec:variation-of-the-metric}.

\subsection{Velocity-field tangent structure and nonlocal metric}
\label{subsec:first-tangent-structure}

We now represent tangent vectors by nonlocal velocity fields.
The potential-space construction gives this representation directly.

For fixed $\rho\in\mathcal P_+(M)$, apply the isometric gradient map
\eqref{eq:isometric-gradient-map} with $J=J_\rho$.

\begin{definition}[Velocity-field tangent space]
\label{def:velocity-field-tangent-space}
The velocity-field tangent representation at $\rho$ is defined by
\begin{equation*}
T_\rho^{\mathrm{vel}}\mathcal P_+(M)
:=
\overline{\nabla}\mathscr H_\rho
=
\overline{
\left\{
\overline{\nabla}\phi:
\phi\in\mathscr D(\mathcal A_\rho)
\right\}
}^{\,
L^2(M\times M,J_\rho)}.
\end{equation*}
\end{definition}

Since $\mathscr H_\rho$ is complete and
$\overline{\nabla}$ is isometric, its image
$T_\rho^{\mathrm{vel}}\mathcal P_+(M)$ is a closed subspace of
$L^2(M\times M,J_\rho)$.

\begin{definition}[Metric in velocity coordinates]
\label{def:velocity-Riemannian-metric}
For $v,w\in T_\rho^{\mathrm{vel}}\mathcal P_+(M)$, define
\begin{equation*}
g_\rho(v,w)
:=
\iint_{M\times M}
v(x,y)w(x,y)\,
J_\rho(\mathrm dx,\mathrm dy).
\end{equation*}
\end{definition}
For potential-generated fields, the metric is the weighted pairing of
their increments, so
\begin{equation}
\label{eq:energy-metric-identity}
\mathcal A_\rho(\phi,\psi)
=
g_\rho
\left(
\overline{\nabla}\phi,
\overline{\nabla}\psi
\right),
\end{equation}
and the corresponding quadratic kinetic energy is
\begin{equation}
\label{eq:quadratic-kinetic-energy}
\mathcal K(\rho,\phi)
:=
\frac12
g_\rho
\left(
\overline{\nabla}\phi,
\overline{\nabla}\phi
\right)
=
\frac12
\|\phi\|_{\mathscr H_\rho}^{2}.
\end{equation}
For $\phi,\psi\in\mathscr D(L_\rho)$, the nonlocal Green formula gives
\begin{equation}
\label{eq:general-integration-by-parts}
\int_M
\phi(x)L_\rho\psi(x)\,\mathrm dx
=
-\mathcal A_\rho(\phi,\psi)
=
-g_\rho
\left(
\overline{\nabla}\phi,
\overline{\nabla}\psi
\right).
\end{equation}


\subsubsection*{Variation of the velocity-field metric}
\label{subsubsec:variation-of-the-metric}

Keep the mobility regularity and smooth density variations of
Section~\ref{subsubsec:density-variation-of-the-dirichlet-form}.
Here the potentials $\xi,\psi$ are held fixed, in contrast to the
density-coordinate variation \eqref{eq:density-metric-variation}.

\begin{lemma}[Variation of the metric]
\label{lem:potential-metric-variation}
For every $\sigma$ as above and every pair of fixed test potentials
$\xi,\psi\in \mathscr H_{\rho}$,
\begin{equation}
\label{eq:potential-metric-variation}
Dg_\rho[\sigma]\left(\overline{\nabla}\xi,\overline{\nabla}\psi\right)
=
2\int_M
\sigma(x)
\Lambda_\rho(\xi,\psi)(x)\,
\mathrm dx.
\end{equation}
\end{lemma}

\begin{proof}
By \eqref{eq:energy-metric-identity} and
Lemma~\ref{lem:density-variation-dirichlet-form}, with $\xi,\psi$ held fixed,
\begin{equation*}
Dg_\rho[\sigma](\Nabla\xi,\Nabla\psi)
=D\mathcal A_\rho[\sigma](\xi,\psi)=2\int_M\sigma\,\Lambda_\rho(\xi,\psi)\,\mathrm dx.
\end{equation*}
\end{proof}


Let $t\mapsto\rho_t$ be a smooth density curve with velocity potential
$\phi_t$, meaning that
\begin{equation}
\label{eq:curve-velocity}
\partial_t\rho_t
=-L_{\rho_t}\phi_t.
\end{equation}
Equivalently,
\begin{equation*}
\partial_t\rho_t
+
\overline{\mathrm{div}}
\left(
J_{\rho_t}\overline{\nabla}\phi_t
\right)
=
0.
\end{equation*}

Since $J_{\rho_t}$ is symmetric, \eqref{eq:curve-velocity} and
Corollary~\ref{cor:integral-LJ-zero} give
\begin{equation*}
\frac{\mathrm d}{\mathrm dt}\int_M\rho_t(x)\,\mathrm dx
=-\int_M(L_{\rho_t}\phi_t)(x)\,\mathrm dx=0.
\end{equation*}
Thus the continuity equation preserves the total mass.

\begin{lemma}[Metric variation along a density curve]
\label{prop:metric-variation-along-curve}
Let $t\mapsto\rho_t\in\mathcal P_+(M)$ be a smooth density curve. Assume that the curve admits a velocity potential $\phi_t\in\mathscr D(L_{\rho_t})$ satisfying \eqref{eq:curve-velocity}.
Let $\xi,\psi\in\mathscr D(\mathcal A_{\rho_t})$ be fixed potentials
with no explicit time dependence. Suppose, in addition, that
$\Lambda_{\rho_t}(\xi,\psi)\in\mathscr D(\mathcal A_{\rho_t})$, so that all the
expressions below are well defined. Then
\begin{align}
\frac{\mathrm d}{\mathrm dt}
g_{\rho_t}\left(\overline{\nabla}\xi,\overline{\nabla}\psi\right)&=2\int_M\partial_t\rho_t(x)\Lambda_{\rho_t}(\xi,\psi)(x)\,\mathrm dx
\label{eq:metric-time-variation}\\
&=2g_{\rho_t}(\overline{\nabla}\Lambda_{\rho_t}(\xi,\psi),\overline{\nabla}\phi_t).
\label{eq:general-metric-variation-final}
\end{align}
\end{lemma}

\begin{proof}
Since $\xi$ and $\psi$ have no explicit time dependence, the time
derivative acts only on the density-dependent metric. Hence, by
Lemma~\ref{lem:potential-metric-variation} and \eqref{eq:curve-velocity},
\begin{align*}
\frac{\mathrm d}{\mathrm dt}
g_{\rho_t}\left(\overline{\nabla}\xi,\overline{\nabla}\psi\right)
&=Dg_{\rho_t}[\partial_t\rho_t]\left(\overline{\nabla}\xi,\overline{\nabla}\psi\right)\\
&=2\int_M\partial_t\rho_t(x)\Lambda_{\rho_t}(\xi,\psi)(x)\,\mathrm dx \\
&=
-2\int_M
\Lambda_{\rho_t}
(\xi,\psi)(x)
L_{\rho_t}\phi_t(x)\,
\mathrm dx.
\end{align*}
The nonlocal Green's formula \eqref{eq:general-integration-by-parts} gives
\begin{align*}
-\int_M\Lambda_{\rho_t}(\xi,\psi)L_{\rho_t}\phi_t\,\mathrm dx
&=g_{\rho_t}(\overline{\nabla}\Lambda_{\rho_t}(\xi,\psi),\overline{\nabla}\phi_t).
\end{align*}
Combining the preceding identities proves
\eqref{eq:general-metric-variation-final}.
\end{proof}

If $\xi_t$ and $\psi_t$ also depend explicitly on time, then the
corresponding product rule reads
\begin{equation*}
\frac{\mathrm d}{\mathrm dt}
g_{\rho_t}
\left(
\overline{\nabla}\xi_t,
\overline{\nabla}\psi_t
\right)
=
g_{\rho_t}
\left(
\overline{\nabla}\partial_t\xi_t,
\overline{\nabla}\psi_t
\right)
\nonumber+
g_{\rho_t}
\left(
\overline{\nabla}\xi_t,
\overline{\nabla}\partial_t\psi_t
\right)
\nonumber+
2g_{\rho_t}
\left(
\overline{\nabla}
\Lambda_{\rho_t}(\xi_t,\psi_t),
\overline{\nabla}\phi_t
\right).
\end{equation*}

\subsubsection*{Equivalence of the two tangent representations}

In the following theorem and proof, the density tangent space is
equipped with its $G_\rho$-completion. The velocity tangent space is
the closed space of Definition~\ref{def:velocity-field-tangent-space}.

\begin{theorem}[Equivalence of the two tangent representations]
\label{thm:tangent-representation-equivalence}
The map
\begin{equation*}
\begin{aligned}
\mathbf I_\rho:\,
T_\rho^{\mathrm{vel}}\mathcal P_+(M)
&\to T_\rho^{\mathrm{dens}}\mathcal P_+(M),\\
\mathbf I_\rho v&:=-\Div(J_\rho v)
\end{aligned}
\end{equation*}
is a linear isometric isomorphism:
\begin{equation*}
G_\rho(\mathbf I_\rho v,\mathbf I_\rho w)
=g_\rho(v,w),
\qquad v,w\in T_\rho^{\mathrm{vel}}\mathcal P_+(M).
\end{equation*}
Its inverse is
\begin{equation*}
\mathbf I_\rho^{-1}(\sigma)
=\Nabla\bigl((-L_\rho)^\dagger\sigma\bigr).
\end{equation*}
\end{theorem}

\begin{proof}
By Definition~\ref{def:velocity-field-tangent-space}, every velocity tangent
vector has the form $\Nabla\phi$ for a potential
$\phi\in\mathscr H_\rho$ modulo constants. It therefore suffices to prove
the theorem for $v=\Nabla\phi$ and $w=\Nabla\psi$.
For every $\xi\in\mathscr H_\rho$, the defining energy-duality identity gives
\[
\begin{aligned}
\langle\xi,\mathbf I_\rho(\Nabla\phi)\rangle_\rho
&=\iint_{M\times M}\Nabla\phi\,\Nabla\xi\,J_\rho(\mathrm dx,\mathrm dy)\\
&=\mathcal A_\rho(\phi,\xi)
=\langle\xi,-L_\rho\phi\rangle_\rho.
\end{aligned}
\]
Thus $\mathbf I_\rho(\Nabla\phi)=-L_\rho\phi$, independently of the potential
representative. Using \eqref{eq:general-potential-pseudoinverse},
\[
\begin{aligned}
\mathbf I_\rho\bigl(\mathbf I_\rho^{-1}(\sigma)\bigr)
&=(-L_\rho)(-L_\rho)^\dagger\sigma=\sigma,\\
\mathbf I_\rho^{-1}\bigl(\mathbf I_\rho(\Nabla\phi)\bigr)
&=\Nabla\bigl[(-L_\rho)^\dagger(-L_\rho)\phi\bigr]
=\Nabla\phi.
\end{aligned}
\]
The pseudoinverse recovers $\phi$ modulo constants, which disappear
under $\Nabla$. Linearity is immediate, and
\[
G_\rho(\mathbf I_\rho(\Nabla\phi),\mathbf I_\rho(\Nabla\psi))
=G_\rho(-L_\rho\phi,-L_\rho\psi)=\mathcal A_\rho(\phi,\psi)
=g_\rho(\Nabla\phi,\Nabla\psi).
\]
\end{proof}
Thus, for $\sigma=-L_\rho\phi$ and $\tau=-L_\rho\psi$, the two metric
representations satisfy
\[
G_\rho(\sigma,\tau)
=g_\rho(\Nabla\phi,\Nabla\psi)
=\mathcal A_\rho(\phi,\psi).
\]


\subsection{Gradients and gradient flows for the nonlocal metric}
\label{subsubsec:nonlocal-Wasserstein-gradient-flows}

We now describe gradients and their flows in the two tangent
representations, starting with density coordinates.

\subsubsection*{Density representation}

For a differentiable functional $\mathcal F$, assume that
$D\mathcal F(\rho)$ on smooth density directions is continuous for
$\|\cdot\|_{G_\rho}$, and extend it to the completion. Let its
cotangent representative, characterized by
$D\mathcal F(\rho)[\sigma]=\langle f_\rho,\sigma\rangle_\rho$, be
\begin{equation*}
  f_\rho:=\frac{\delta\mathcal F}{\delta\rho}(\rho)\in\mathscr H_\rho.
\end{equation*}
The defining property of the density gradient and the flat map give,
for every density tangent vector $\sigma$,
\[
\left\langle
\bigl(\mathrm{grad}^{\mathrm{dens}}_\rho\mathcal F\bigr)^\flat,
\sigma\right\rangle_\rho
=G_\rho(\mathrm{grad}^{\mathrm{dens}}_\rho\mathcal F,\sigma)=D\mathcal F(\rho)[\sigma]
=\langle f_\rho,\sigma\rangle_\rho.
\]
Hence $\bigl(\mathrm{grad}^{\mathrm{dens}}_\rho\mathcal F\bigr)^\flat=f_\rho$.
Applying the sharp map and \eqref{eq:metric-Legendre-maps} yields
\begin{equation}
\label{eq:Riemannian-gradient}
\mathrm{grad}^{\mathrm{dens}}_\rho\mathcal F
=f_\rho^\sharp
=-L_\rho f_\rho
=\mathbf I_\rho(\Nabla f_\rho).
\end{equation}

Consequently, the negative gradient-flow equation in the completed
density space is $\partial_t\rho_t=L_{\rho_t}f_{\rho_t}$.
If $f_{\rho_t}\in\mathscr D(L_{\rho_t})$ and $L_{\rho_t}f_{\rho_t}$
represents the smooth curve velocity, the equation takes the following
form for almost every $x$,
\begin{equation}
\label{eq:density-gradient-flow}
\begin{aligned}
\partial_t\rho_t(x)
&=-\bigl(\mathrm{grad}^{\mathrm{dens}}_{\rho_t}\mathcal F\bigr)(x)
=(L_{\rho_t}f_{\rho_t})(x)\\
&=2 \mathrm{p.v.} \int_M
\bigl[f_{\rho_t}(y)-f_{\rho_t}(x)\bigr]
\theta_{\rho_t}(x,y)K(x,y)\,\mathrm dy.
\end{aligned}
\end{equation}
The equation is understood through the Dirichlet-form pairing.

\subsubsection*{Velocity-field representation}

A velocity field
\[
\overline{\nabla}\phi
\in
T_\rho^{\mathrm{vel}}\mathcal P_+(M)
\]
induces a density tangent vector (a smooth zero-mass variation only when the image
defined through energy duality has a smooth representative)
\begin{equation*}
\sigma_\phi
:=
-L_\rho\phi.
\end{equation*}

Let $\mathcal F:\mathcal P_+(M)\to\mathbb R$ be a differentiable functional.
Its gradient on $T_\rho^{\mathrm{vel}}\mathcal P_+(M)$ is defined by
\begin{equation*}
g_\rho(\mathrm{grad}_{\rho}^{\mathrm{vel}}\mathcal{F},\overline{\nabla}\phi)
=
D\mathcal F(\rho)[\sigma_\phi].
\end{equation*}

\begin{proposition}[Wasserstein gradient for the nonlocal metric]
\label{thm:nonlocal-Wasserstein-gradient}
Let $\mathcal F:\mathcal P_+(M)\to\mathbb R$ be differentiable at $\rho$, with $D\mathcal F(\rho)$ continuous for $\|\cdot\|_{G_\rho}$
and represented in energy duality by
$f_\rho=\frac{\delta\mathcal F}{\delta\rho}(\rho)\in\mathscr H_\rho$.
Then the gradient of $\mathcal F$ with respect to $g_\rho$ in the
velocity-field representation is
\begin{equation*}
\mathrm{grad}_{\rho}^{\mathrm{vel}}\mathcal F=\overline{\nabla}
\frac{\delta\mathcal F}{\delta\rho}.
\end{equation*}
\end{proposition}

\begin{proof}
For every
$\phi\in\mathscr H_\rho$ and $\sigma_\phi=-L_\rho\phi$,
\[
D\mathcal F(\rho)[\sigma_\phi]
=\langle f_\rho,\sigma_\phi\rangle_\rho
=\mathcal A_\rho(f_\rho,\phi)
=g_\rho(\Nabla f_\rho,\Nabla\phi).
\]
Then, the nonlocal Green's formula \eqref{eq:general-integration-by-parts} yields
\[
-\int_M
f_\rho L_\rho\phi\,\mathrm dx
=
g_\rho
\left(
\overline{\nabla}f_\rho,
\overline{\nabla}\phi
\right).
\]
Therefore,
\[
g_\rho
\left(
\mathrm{grad}_{\rho}^{\mathrm{vel}}\mathcal F,
\overline{\nabla}\phi
\right)
=
g_\rho
\left(
\overline{\nabla}f_\rho,
\overline{\nabla}\phi
\right)
\]
for every tangent velocity field $\overline{\nabla}\phi$. The nondegeneracy of $g_\rho$
gives
\[
\mathrm{grad}_{\rho}^{\mathrm{vel}}\mathcal F=\overline{\nabla}f_\rho=\overline{\nabla}\frac{\delta\mathcal F}{\delta\rho}.
\]
\end{proof}

The negative gradient flow is obtained by choosing the velocity
\(
v_t
=-\mathrm{grad}_{\rho}^{\mathrm{vel}}\mathcal F
\)
in the nonlocal continuity equation
\(
\partial_t\rho_t
+
\overline{\mathrm{div}}
\bigl(
J_{\rho_t}v_t
\bigr)
=
0.
\)
Consequently,
\begin{equation*}
\partial_t\rho_t
=L_{\rho_t}\frac{\delta\mathcal F}{\delta\rho}(\rho_t).
\end{equation*}

By \eqref{eq:Riemannian-gradient},
$\mathbf I_\rho(\mathrm{grad}^{\mathrm{vel}}_\rho\mathcal F)
=\mathrm{grad}^{\mathrm{dens}}_\rho\mathcal F$.
Hence the velocity-field and density formulations describe the same
negative gradient flow.

\subsubsection*{Energy dissipation}

\begin{proposition}[Dissipation along a negative gradient flow]
\label{prop:gradient-flow-dissipation}
Let $\rho_t$ be a smooth density curve satisfying
$\partial_t\rho_t=L_{\rho_t}f_{\rho_t}$ in the completed density
space. Assume that $\mathcal F$ is differentiable along the curve,
with $D\mathcal F(\rho_t)$ continuous for $\|\cdot\|_{G_{\rho_t}}$
and represented by $f_{\rho_t}\in\mathscr H_{\rho_t}$.
Then
\begin{equation}
\label{eq:examples-entropy-dissipation}
\begin{aligned}
\frac{\mathrm d}{\mathrm dt}\mathcal F(\rho_t)
&=-G_{\rho_t}(\partial_t\rho_t,\partial_t\rho_t)
=-g_{\rho_t}(\Nabla f_{\rho_t},\Nabla f_{\rho_t})
\leq0.
\end{aligned}
\end{equation}
\end{proposition}

\begin{proof}
The chain rule, \eqref{eq:density-gradient-flow}, and the nonlocal
Green's formula give
\[
\frac{\mathrm d}{\mathrm dt}\mathcal F(\rho_t)
=D\mathcal F(\rho_t)[\partial_t\rho_t]
=\langle f_{\rho_t},\partial_t\rho_t\rangle_{\rho_t}=\int_Mf_{\rho_t}\,\partial_t\rho_t\,\mathrm dx
=\int_Mf_{\rho_t}L_{\rho_t}f_{\rho_t}\,\mathrm dx
=-\mathcal A_{\rho_t}(f_{\rho_t},f_{\rho_t}).
\]
Since $\partial_t\rho_t=L_{\rho_t}f_{\rho_t}$ is represented by
the potential $-f_{\rho_t}$, the definitions of $G_\rho$ and
$g_\rho$ yield the remaining equalities.
\end{proof}

Entropy gradient flows for the logarithmic and geometric means are
discussed in Section~\ref{subsec:logarithmic-mean-fractional-heat-revised}.

\section{Covariant calculus and Hamiltonian flows}
\label{sec:covariant-calculus}

For the nonlocal metric on Wasserstein space, we first construct the
Levi-Civita connection and covariant differentiation in density coordinates.
The tangent-space isometry then transports these
operations to velocity fields. We conclude with the action principle and
the nonlocal Hamilton--Jacobi and nonlocal Newton formulations of Hamiltonian flows.

\begin{assumption}\label{ass:Lambda-domain}
Throughout this section, in addition to Assumption~\ref{ass:reference-kernel} and Assumption~\ref{ass:smooth-realization}, we assume that $\Lambda_\rho(\phi,\psi)\in \mathscr D(L_\rho)$ for every smooth potentials $\phi,\psi$.
\end{assumption}

\subsection{Levi-Civita covariant differentiation in density coordinates}
\label{subsec:density-covariant-calculus}

The density-metric variation \eqref{eq:density-metric-variation}
determines the connection through the Koszul formula, with the
density-variation operator $\Lambda_\rho$ entering the Christoffel term.

\begin{theorem}[Levi-Civita Christoffel term]
\label{thm:connection}
For smooth $\sigma=-L_\rho\phi$ and $\tau=-L_\rho\psi$, the Christoffel term
in affine density coordinates is
\begin{equation}
\label{eq:Christoffel-density}
\Gamma_W(\sigma,\tau)
=\frac12(DL_\rho[\sigma])\psi
+\frac12(DL_\rho[\tau])\phi
-L_\rho\Lambda_\rho(\phi,\psi).
\end{equation}
Thus, for a density vector field $\tau=\tau(\rho)$, the
Levi-Civita connection is
\begin{equation}
\label{eq:full-connection}
\nabla_\sigma\tau
=D\tau(\rho)[\sigma]+\Gamma_W(\sigma,\tau).
\end{equation}
\end{theorem}

\begin{proof}
The right-hand side of \eqref{eq:Christoffel-density} is smooth
by Assumption~\ref{ass:smooth-realization}. Each term has zero
integral: this follows from mass conservation for $L_\rho$, and
from differentiating $\int_M L_\rho\psi\,\mathrm dx=0$ for the
operator-variation terms. It therefore defines a smooth density
tangent vector, depending smoothly and bilinearly on $\sigma,\tau$. For constant density vector fields in affine coordinates, the Koszul
formula reads
\begin{equation}
\label{eq:koszul}
2G_\rho(\Gamma_W(\sigma,\tau),\eta)
=DG_\rho[\sigma](\tau,\eta)
+DG_\rho[\tau](\sigma,\eta)
-DG_\rho[\eta](\sigma,\tau).
\end{equation}
Writing $\eta=-L_\rho\xi$, substituting
\eqref{eq:density-metric-variation}, and dividing by two, we obtain
\begin{equation}
\label{eq:koszul-expanded}
G_\rho(\Gamma_W(\sigma,\tau),\eta)
=-\int_M\sigma\,\Lambda_\rho(\psi,\xi)\,\mathrm dx
-\int_M\tau\,\Lambda_\rho(\phi,\xi)\,\mathrm dx+\int_M\eta\,\Lambda_\rho(\phi,\psi)\,\mathrm dx.
\end{equation}
To express these terms as pairings with $\xi$, apply
\eqref{eq:key-duality} separately:
\[
\begin{aligned}
-\int_M\sigma\,\Lambda_\rho(\psi,\xi)\,\mathrm dx
&=\frac12\int_M\xi(DL_\rho[\sigma])\psi\,\mathrm dx,\\
-\int_M\tau\,\Lambda_\rho(\phi,\xi)\,\mathrm dx
&=\frac12\int_M\xi(DL_\rho[\tau])\phi\,\mathrm dx.
\end{aligned}
\]
For the third term, $\eta=-L_\rho\xi$ and symmetry of $L_\rho$
give
\[
\int_M\eta\,\Lambda_\rho(\phi,\psi)\,\mathrm dx=-\int_M(L_\rho\xi)\Lambda_\rho(\phi,\psi)\,\mathrm dx=-\int_M\xi L_\rho\Lambda_\rho(\phi,\psi)\,\mathrm dx.
\]
Substituting these identities into \eqref{eq:koszul-expanded} yields
\[
G_\rho(\Gamma_W(\sigma,\tau),\eta)
=\int_M\xi\left[
\frac12(DL_\rho[\sigma])\psi
+\frac12(DL_\rho[\tau])\phi
-L_\rho\Lambda_\rho(\phi,\psi)
\right]\,\mathrm dx.
\]
Since
$G_\rho(\Gamma_W(\sigma,\tau),\eta)
=\int_M\Gamma_W(\sigma,\tau)\xi\,\mathrm dx$ by
\eqref{eq:density-duality}, nondegeneracy of $G_\rho$ yields
\eqref{eq:Christoffel-density}. Adding the ordinary derivative
$D\tau(\rho)[\sigma]$ gives \eqref{eq:full-connection} for a general
vector field.
The Christoffel term is symmetric in $\sigma,\tau$, and the
Koszul construction gives metric compatibility.
\end{proof}

Substituting the potentials $\phi=-L_\rho^\dagger\sigma$ and
$\psi=-L_\rho^\dagger\tau$ into \eqref{eq:Christoffel-density} gives, by the bilinearity of $\Lambda_\rho$,
\begin{equation*}
\begin{aligned}
\Gamma_W(\sigma,\tau)
&=-\frac12(DL_\rho[\sigma])L_\rho^\dagger\tau
-\frac12(DL_\rho[\tau])L_\rho^\dagger\sigma-L_\rho\Lambda_\rho
(L_\rho^\dagger\sigma,L_\rho^\dagger\tau).
\end{aligned}
\end{equation*}

\subsubsection*{Covariant differentiation along a density curve}

For a density tangent field $\tau_t$ along $\rho_t$, the
Levi-Civita connection \eqref{eq:full-connection} induces
\begin{equation}
\label{eq:covariant-definition-density}
\frac{D}{\mathrm dt}\tau_t
:=\partial_t\tau_t+\Gamma_W(\partial_t\rho_t,\tau_t).
\end{equation}
In particular, if $\tau_t=\tau(\rho_t)$, this is
$\nabla_{\partial_t\rho_t}\tau$.

All formulas below are understood on smooth fields in the relevant domains, with
$DL_\rho[\sigma]$ differentiating the operator at a fixed potential.

\begin{corollary}[Covariant derivative in density coordinates]
\label{prop:density-covariant}
Let $\rho_t$ be a smooth density curve and $\tau_t$ a tangent field
along it, represented by potentials $\phi_t,\psi_t$ through
\begin{equation}
\label{eq:sigma-along-curve}
\partial_t\rho_t=-L_{\rho_t}\phi_t,
\qquad
\tau_t=-L_{\rho_t}\psi_t.
\end{equation}
Then
\begin{equation}
\label{eq:density-covariant}
\frac{D}{\mathrm dt}\tau_t
=-\frac12(DL_{\rho_t}[\partial_t\rho_t])\psi_t
+\frac12(DL_{\rho_t}[\tau_t])\phi_t -L_{\rho_t}
\bigl(\partial_t\psi_t+\Lambda_{\rho_t}(\phi_t,\psi_t)\bigr).
\end{equation}
\end{corollary}

\begin{proof}
Differentiating the quantity $L_{\rho_t}\psi_t$ gives
\begin{equation}
\label{eq:ordinary-derivative}
\partial_t\tau_t
=-(DL_{\rho_t}[\partial_t\rho_t])\psi_t
-L_{\rho_t}\partial_t\psi_t.
\end{equation}
Substituting $\partial_t\rho_t=-L_{\rho_t}\phi_t$ and
$\tau_t=-L_{\rho_t}\psi_t$ into \eqref{eq:Christoffel-density},
\[
\Gamma_W(\partial_t\rho_t,\tau_t)
=\frac12(DL_{\rho_t}[\partial_t\rho_t])\psi_t
+\frac12(DL_{\rho_t}[\tau_t])\phi_t -L_{\rho_t}\Lambda_{\rho_t}(\phi_t,\psi_t).
\]
Adding this to \eqref{eq:ordinary-derivative} gives
\[
\frac{D}{\mathrm dt}\tau_t
=-(DL_{\rho_t}[\partial_t\rho_t])\psi_t
-L_{\rho_t}\partial_t\psi_t+\frac12(DL_{\rho_t}[\partial_t\rho_t])\psi_t
+\frac12(DL_{\rho_t}[\tau_t])\phi_t -L_{\rho_t}\Lambda_{\rho_t}(\phi_t,\psi_t).
\]
Combining the two terms containing
$(DL_{\rho_t}[\partial_t\rho_t])\psi_t$ yields
\eqref{eq:density-covariant}.
\end{proof}

\subsubsection*{Metric compatibility}

The Levi-Civita property gives, for density tangent fields
$\eta_t,\tau_t$,
\begin{equation}
\label{eq:density-metric-compatibility}
\frac{\mathrm d}{\mathrm dt}G_{\rho_t}(\eta_t,\tau_t)
=G_{\rho_t}\left(\frac{D}{\mathrm dt}\eta_t,\tau_t\right)
+G_{\rho_t}\left(\eta_t,\frac{D}{\mathrm dt}\tau_t\right).
\end{equation}
Indeed, for fixed density tangent vectors, the Koszul formula
\eqref{eq:koszul} gives
\[
DG_\rho[\sigma](\eta,\tau)
=G_\rho(\Gamma_W(\sigma,\eta),\tau)
+G_\rho(\eta,\Gamma_W(\sigma,\tau)).
\]
The ordinary chain rule along $\rho_t$, together with
\eqref{eq:covariant-definition-density}, then yields
\eqref{eq:density-metric-compatibility}: the Christoffel terms account for
the dependence of the metric on $\rho_t$.
In particular, the inner product of parallel density fields is constant.
We will recover the same property in velocity-field coordinates below.

\subsection{Levi-Civita covariant differentiation in velocity-field coordinates}
\label{subsec:velocity-covariant-calculus}

The isometry $\mathbf I_\rho$ transports the density connection to
velocity fields. For velocity-field sections $v,w$, set
\begin{equation}
\label{eq:velocity-Levi-Civita-connection}
\nabla^{\mathrm{vel}}_v w
:=\mathbf I_\rho^{-1}
\bigl(\nabla_{\mathbf I_\rho v}(\mathbf I w)\bigr),
\end{equation}
where $(\mathbf I w)(\rho)=\mathbf I_\rho w(\rho)$ is a density
vector field. The derivative on the right therefore acts on both
$\mathbf I_\rho$ and $w(\rho)$. With the Lie bracket transported by
$\mathbf I$, this connection is torsion-free and compatible with $g_\rho$, since the density connection
has these properties and $\mathbf I_\rho$ is an isometry.
Along a density curve, we denote the covariant derivative induced by
this connection by $D_t$.

\begin{proposition}[Equivalence of the two covariant derivatives]
\label{thm:equivalence}
Let $\tau_t=-L_{\rho_t}\psi_t$ and $\eta_t=-L_{\rho_t}\xi_t$.
Then
\begin{equation}
\label{eq:pushforward-covariant}
\frac{D}{\mathrm dt}\tau_t
=\mathbf I_{\rho_t}(D_t\Nabla\psi_t),
\end{equation}
and
\begin{equation}
\label{eq:dual-equivalence}
G_{\rho_t}\left(\frac{D}{\mathrm dt}\tau_t,\eta_t\right)
=g_{\rho_t}(D_t\Nabla\psi_t,\Nabla\xi_t).
\end{equation}
\end{proposition}

\begin{proof}
Since $\tau_t=\mathbf I_{\rho_t}(\Nabla\psi_t)$, the transported
connection \eqref{eq:velocity-Levi-Civita-connection} gives
\begin{equation}
\label{eq:covariant-identification-along-curve}
D_t\Nabla\psi_t
=\mathbf I_{\rho_t}^{-1}
\left(\frac{D}{\mathrm dt}\tau_t\right).
\end{equation}
Applying $\mathbf I_{\rho_t}$ to
\eqref{eq:covariant-identification-along-curve} gives
\eqref{eq:pushforward-covariant}. Since
$\eta_t=\mathbf I_{\rho_t}(\Nabla\xi_t)$, the isometry yields
\[
G_{\rho_t}\left(\frac{D}{\mathrm dt}\tau_t,\eta_t\right)
=G_{\rho_t}\bigl(\mathbf I_{\rho_t}(D_t\Nabla\psi_t),
\mathbf I_{\rho_t}(\Nabla\xi_t)\bigr)=g_{\rho_t}(D_t\Nabla\psi_t,\Nabla\xi_t).
\]
\end{proof}
Thus, parallel transport agrees in both representations:
\begin{equation*}
\frac{D}{\mathrm dt}\tau_t=0
\quad\Longleftrightarrow\quad
D_t\Nabla\psi_t=0.
\end{equation*}

\subsubsection*{Explicit potential representation}

Recall the representations \eqref{eq:sigma-along-curve},
$\partial_t\rho_t=-L_{\rho_t}\phi_t$ and $\tau_t=-L_{\rho_t}\psi_t$.
Applying $\mathbf I_{\rho_t}^{-1}$ to \eqref{eq:density-covariant} gives
\begin{equation}
\label{eq:general-covariant-potential}
D_t\Nabla\psi_t
=\Nabla\Bigl[\partial_t\psi_t
+\Lambda_{\rho_t}(\phi_t,\psi_t)+\frac12L_{\rho_t}^\dagger
\bigl((DL_{\rho_t}[\partial_t\rho_t])\psi_t
-(DL_{\rho_t}[\tau_t])\phi_t\bigr)\Bigr].
\end{equation}

\begin{corollary}[Potential formulation of ovariant differentiation]
\label{thm:weak-potential}
Let $\rho_t$ be a smooth density curve with velocity potential $\phi_t$,
so that
\[
\partial_t\rho_t=-L_{\rho_t}\phi_t.
\]
For any smooth time-dependent velocity field $\Nabla\psi_t$ along
$\rho_t$ and test potential $\xi_t$, the following identity holds
pointwise in time:
\begin{equation}
\label{eq:weak-potential}
\begin{aligned}
g_{\rho_t}(D_t\Nabla\psi_t,\Nabla\xi_t)
&=g_{\rho_t}(\Nabla\partial_t\psi_t,\Nabla\xi_t)+g_{\rho_t}
(\Nabla\Lambda_{\rho_t}(\psi_t,\xi_t),\Nabla\phi_t)\\
&\quad+g_{\rho_t}
(\Nabla\Lambda_{\rho_t}(\phi_t,\psi_t),\Nabla\xi_t)-g_{\rho_t}
(\Nabla\Lambda_{\rho_t}(\phi_t,\xi_t),\Nabla\psi_t).
\end{aligned}
\end{equation}
\end{corollary}

\begin{proof}
Let $\tau_t=-L_{\rho_t}\psi_t$. By \eqref{eq:dual-equivalence},
\begin{equation}
\label{eq:covariant-duality-along-curve}
g_{\rho_t}(D_t\Nabla\psi_t,\Nabla\xi_t)
=\int_M\xi_t\frac{D}{\mathrm dt}\tau_t\,\mathrm dx.
\end{equation}
Insert \eqref{eq:density-covariant} into
\eqref{eq:covariant-duality-along-curve} and treat its three
contributions separately. By \eqref{eq:key-duality},
$\partial_t\rho_t=-L_{\rho_t}\phi_t$, and nonlocal Green's formula,
\[
\begin{aligned}
-\frac12\int_M\xi_t(DL_{\rho_t}[\partial_t\rho_t])\psi_t\,\mathrm dx
&=\int_M\partial_t\rho_t\,\Lambda_{\rho_t}(\psi_t,\xi_t)\,\mathrm dx\\
&=-\int_M(L_{\rho_t}\phi_t)\Lambda_{\rho_t}(\psi_t,\xi_t)\,\mathrm dx\\
&=g_{\rho_t}
(\Nabla\Lambda_{\rho_t}(\psi_t,\xi_t),\Nabla\phi_t).
\end{aligned}
\]
Similarly, using
$\tau_t=-L_{\rho_t}\psi_t$,
\[
\frac12\int_M\xi_t(DL_{\rho_t}[\tau_t])\phi_t\,\mathrm dx
=-g_{\rho_t}
(\Nabla\Lambda_{\rho_t}(\phi_t,\xi_t),\Nabla\psi_t).
\]
Finally,
\[
-\int_M\xi_t L_{\rho_t}
\bigl(\partial_t\psi_t+\Lambda_{\rho_t}(\phi_t,\psi_t)\bigr)\,
\mathrm dx
=g_{\rho_t}(\Nabla\partial_t\psi_t,\Nabla\xi_t)
+g_{\rho_t}(\Nabla\Lambda_{\rho_t}(\phi_t,\psi_t),\Nabla\xi_t).
\]
Adding the three identities proves \eqref{eq:weak-potential}.
\end{proof}

\subsubsection*{Metric compatibility}

\begin{corollary}[Metric compatibility along curves]
\label{prop:metric-compatibility}
For smooth potential fields $\xi_t,\psi_t$ along $\rho_t$,
\begin{equation*}
\frac{\mathrm d}{\mathrm dt}
g_{\rho_t}(\Nabla\xi_t,\Nabla\psi_t)
=g_{\rho_t}(D_t\Nabla\xi_t,\Nabla\psi_t)+g_{\rho_t}(\Nabla\xi_t,D_t\Nabla\psi_t).
\end{equation*}
\end{corollary}

\begin{proof}
The ordinary product rule gives
\[
\frac{\mathrm d}{\mathrm dt}g_{\rho_t}(\Nabla\xi_t,\Nabla\psi_t)
=g_{\rho_t}(\Nabla\partial_t\xi_t,\Nabla\psi_t)
+g_{\rho_t}(\Nabla\xi_t,\Nabla\partial_t\psi_t)
+Dg_{\rho_t}[\partial_t\rho_t](\Nabla\xi_t,\Nabla\psi_t),
\]
where the potentials are held fixed in the last term.
By \eqref{eq:potential-metric-variation},
\[
Dg_{\rho_t}[\partial_t\rho_t](\Nabla\xi_t,\Nabla\psi_t)
=2\int_M\partial_t\rho_t\,
\Lambda_{\rho_t}(\xi_t,\psi_t)\,\mathrm dx=2g_{\rho_t}
(\Nabla\Lambda_{\rho_t}(\xi_t,\psi_t),\Nabla\phi_t).
\]
Now add \eqref{eq:weak-potential} applied to the ordered pairs of
differentiated and test potentials $(\xi_t,\psi_t)$ and $(\psi_t,\xi_t)$.
The terms containing
$\Lambda_{\rho_t}(\phi_t,\xi_t)$ and
$\Lambda_{\rho_t}(\phi_t,\psi_t)$ cancel with their negatives,
leaving
\[
\begin{aligned}
&g_{\rho_t}(D_t\Nabla\xi_t,\Nabla\psi_t)
+g_{\rho_t}(\Nabla\xi_t,D_t\Nabla\psi_t)\\
&=g_{\rho_t}(\Nabla\partial_t\xi_t,\Nabla\psi_t)
+g_{\rho_t}(\Nabla\xi_t,\Nabla\partial_t\psi_t)+2g_{\rho_t}
(\Nabla\Lambda_{\rho_t}(\xi_t,\psi_t),\Nabla\phi_t).
\end{aligned}
\]
This is exactly the ordinary product rule above.
\end{proof}
Consequently, the inner product of two parallel fields, and in
particular the norm of a parallel field, is constant along the curve.

The metric-variation identity \eqref{eq:metric-time-variation} fixes
only the sum of the two connection pairings. The last two terms in
\eqref{eq:weak-potential} cancel when $\xi_t$ and $\psi_t$ are
interchanged; they are determined by the torsion-free condition through
the Koszul formula, not by metric variation alone.

\subsection{Wasserstein Hamiltonian flows in the two tangent representations}
\label{subsec:Hamiltonian-flows-two-representations}

We first formulate the action principle and its Lagrangian counterpart for a
general Hamiltonian. The common metric supplies the quadratic
kinetic energy; we derive its Hamiltonian flow from the action principle and express
the dynamics through the density and velocity-field covariant derivatives. We then differentiate gradient flows in time to obtain their Hamiltonian lifts.

\subsubsection{Hamiltonian variational principle and Legendre duality}

For the action formulation, assume that $K$ is irreducible.
Let $\mathcal H(\rho,\phi)$ be a differentiable Hamiltonian on the
cotangent phase space of $\mathcal P_+(M)$, with smooth variational
derivatives. Its potential variable belongs to $\mathscr H_\rho$,
the energy completion of $\mathscr D(\mathcal A_\rho)/\mathbb R$.
Such a Hamiltonian may be induced by a differentiable functional on representatives
$\phi\in\mathscr D(\mathcal A_\rho)$, denoted by the same symbol, satisfying
\[
\mathcal H(\rho,\phi+c)=\mathcal H(\rho,\phi),
\qquad c\in\mathbb R.
\]
This invariance makes the functional well defined on
$\mathscr D(\mathcal A_\rho)/\mathbb R$; its extension to $\mathscr H_\rho$ is
assumed whenever the completion is used.
Here $\phi$ is the conjugate potential; for a general Hamiltonian it
need not be the velocity potential of the density curve.
Differentiating the representative functional in the constant direction
$1$ gives
\[
0=\left.\frac{\mathrm d}{\mathrm dc}\right|_{c=0}
\mathcal H(\rho,\phi+c)
=D_\phi\mathcal H(\rho,\phi)[1]
=\int_M\frac{\delta\mathcal H}{\delta\phi}(\rho,\phi)\,\mathrm dx.
\]
Hence $\delta\mathcal H/\delta\phi$ has zero mass and belongs to the
density tangent space.

For a smooth density trajectory $\rho_\cdot$ and a smooth conjugate potential $\phi_\cdot$, define
\begin{equation}
\label{eq:canonical-action-functional}
\mathscr A[\rho_\cdot,\phi_\cdot]
:=
\int_0^T
\left[
\int_M\phi_t(x)\partial_t\rho_t(x)\,\mathrm dx
-\mathcal H(\rho_t,\phi_t)
\right]\,\mathrm dt.
\end{equation}
The canonical pairing term changes by
$c(t)\int_M\partial_t\rho_t\,\mathrm dx=0$ under
$\phi_t\mapsto\phi_t+c(t)$, while the Hamiltonian is unchanged by
the quotient property above. Thus the action is independent of the
potential representative.

\begin{lemma}[Hamiltonian variational principle]
\label{thm:Hamiltonian-variational-principle}
A smooth trajectory $(\rho_\cdot,\phi_\cdot)$ is stationary for
\eqref{eq:canonical-action-functional} under mass-preserving density
variations with fixed density endpoints and unrestricted potential
variations if and only if
\begin{equation}
\label{eq:canonical-Hamiltonian-system}
\begin{cases}
\displaystyle
\partial_t\rho_t
=\frac{\delta\mathcal H}{\delta\phi}(\rho_t,\phi_t),\\[0.8em]
\displaystyle
\partial_t\phi_t
+\frac{\delta\mathcal H}{\delta\rho}(\rho_t,\phi_t)=C(t),
\end{cases}
\end{equation}
where $C(t)$ is spatially constant.
\end{lemma}

\begin{proof}
Let $\sigma_t,\psi_t$ be the density and potential variations, with
$\int_M\sigma_t\,\mathrm dx=0$ and $\sigma_0=\sigma_T=0$.
The first variation is
\[
\begin{aligned}
\delta\mathscr A
&=\int_0^T\int_M
\left[
\psi_t\partial_t\rho_t+\phi_t\partial_t\sigma_t
-\frac{\delta\mathcal H}{\delta\rho}\sigma_t
-\frac{\delta\mathcal H}{\delta\phi}\psi_t
\right]\,\mathrm dx\,\mathrm dt\\
&=\int_0^T\int_M
\left[
\psi_t\left(\partial_t\rho_t-\frac{\delta\mathcal H}{\delta\phi}\right)
-\sigma_t\left(\partial_t\phi_t+\frac{\delta\mathcal H}{\delta\rho}\right)
\right]\,\mathrm dx\,\mathrm dt.
\end{aligned}
\]
Here the derivatives of $\mathcal H$ are evaluated at $(\rho_t,\phi_t)$,
and the second equality uses integration by parts in time.
Arbitrariness of $\psi_t$ gives the first Hamilton equation.
Since $\sigma_t$ ranges over zero-mass variations, its coefficient is
spatially constant, giving the second equation.
Conversely, \eqref{eq:canonical-Hamiltonian-system} makes the displayed
variation vanish: the first coefficient is zero and the second pairs
to zero with $\sigma_t$.
\end{proof}

\paragraph{Legendre transform and Euler--Lagrange equation.}
\label{rem:Hamiltonian-Lagrangian-duality}

For the fiberwise transform, the potential and density variables
are paired by energy duality on the completed spaces. The
Euler--Lagrange calculation below is restricted to smooth
trajectories whose maximizing potential and variational derivatives
have smooth representatives. Suppose that $\mathcal H(\rho,\cdot)$ is strictly convex on potential
classes and that the fiberwise supremum
\begin{equation*}
\mathcal L(\rho,\sigma)
:=
\sup_{\phi\in T_\rho^*\mathcal P_+(M)}
\left\{\langle\phi,\sigma\rangle_\rho-\mathcal H(\rho,\phi)\right\},
\qquad \sigma\in T_\rho^{\mathrm{dens}}\mathcal P_+(M),
\end{equation*}
is attained at a potential $\phi=\phi(\rho,\sigma)$ depending smoothly
on the variables under consideration. The derivative identities below are taken along smooth density and tangent variations. The resulting \emph{Legendre transform} is
characterized by
\begin{equation}
\label{eq:Hamiltonian-Legendre-relation}
\sigma=\frac{\delta\mathcal H}{\delta\phi}(\rho,\phi),
\qquad
\frac{\delta\mathcal L}{\delta\sigma}(\rho,\sigma)=\phi,
\qquad
\frac{\delta\mathcal L}{\delta\rho}(\rho,\sigma)
=-\frac{\delta\mathcal H}{\delta\rho}(\rho,\phi).
\end{equation}

The corresponding configuration-space action is
\begin{equation*}
\mathscr S[\rho_\cdot]
:=\int_0^T\mathcal L(\rho_t,\partial_t\rho_t)\,\mathrm dt.
\end{equation*}
Stationarity is equivalent to the constrained
Euler--Lagrange equation
\begin{equation}
\label{eq:general-Euler-Lagrange}
\partial_t\left[
\frac{\delta\mathcal L}{\delta\sigma}(\rho_t,\partial_t\rho_t)
\right]
-\frac{\delta\mathcal L}{\delta\rho}(\rho_t,\partial_t\rho_t)
=C(t).
\end{equation}
Identities \eqref{eq:Hamiltonian-Legendre-relation}
identify this
equation with \eqref{eq:canonical-Hamiltonian-system}.
Along a Legendre-related pair, $\mathscr S[\rho_\cdot]
=\mathscr A[\rho_\cdot,\phi_\cdot]$.

\subsubsection{Mechanical Hamiltonians and their Lagrangians}

For the quadratic kinetic energy induced by the nonlocal metric, consider
the mechanical Hamiltonian
\begin{equation*}
\mathcal H_0(\rho,\phi)
:=
\mathcal K(\rho,\phi)
+
\mathcal F(\rho).
\end{equation*}
Corollary~\ref{cor:kinetic-density-variation} expresses the
kinetic-energy variations through the Dirichlet form and the
density-variation operator:
\begin{equation*}
\begin{aligned}
D_\phi\mathcal K(\rho,\phi)[\psi]
&=
\mathcal A_\rho(\phi,\psi)
=
-\int_M
\psi L_\rho\phi\,\mathrm dx,
\\
D_\rho\mathcal K(\rho,\phi)[\sigma]
&=
\int_M
\sigma(x)
\Lambda_\rho(\phi,\phi)(x)\,
\mathrm dx.
\end{aligned}
\end{equation*}
Consequently,
\begin{equation}
\label{eq:Hamiltonian-variational-derivatives}
\frac{\delta\mathcal H_0}{\delta\phi}
=
-L_\rho\phi,
\qquad
\frac{\delta\mathcal H_0}{\delta\rho}
=
\Lambda_\rho(\phi,\phi)
+
\frac{\delta\mathcal F}{\delta\rho}.
\end{equation}

Let $\mathscr A_0$ denote the action
\eqref{eq:canonical-action-functional} with $\mathcal H=\mathcal H_0$.
Applying Lemma~\ref{thm:Hamiltonian-variational-principle} to
\eqref{eq:Hamiltonian-variational-derivatives} gives the following system.

\begin{theorem}[Mechanical Hamiltonian equations]
\label{thm:quadratic-Hamiltonian-equations}
Let $(\rho_t,\phi_t)$ be a smooth stationary point of
$\mathscr A_0$ under variations preserving the total mass of
$\rho_t$, with fixed density endpoints. Then
\begin{equation}
\label{eq:general-Hamiltonian-system}
\begin{cases}
\displaystyle
\partial_t\rho_t
=
-L_{\rho_t}\phi_t,
\\[0.8em]
\displaystyle
\partial_t\phi_t
+
\Lambda_{\rho_t}
(\phi_t,\phi_t)
+
\dfrac{\delta\mathcal F}{\delta\rho}(\rho_t)
=
C(t),
\end{cases}
\end{equation}
where $C(t)$ is independent of the spatial variable.
\end{theorem}

The second equation is the nonlocal Hamilton--Jacobi equation, while the first
can equivalently be written as
\begin{equation*}
\partial_t\rho_t
+
\overline{\mathrm{div}}
\left(
J_{\rho_t}\overline{\nabla}\phi_t
\right)
=
0.
\end{equation*}
Thus the conjugate potential is also the velocity potential of $\rho_t$;
we use $\phi_t$ for both in the mechanical dynamics below.

Replacing $\phi_t$ by $\phi_t-\int_0^t C(s)\,\mathrm ds$ sets
$C(t)=0$ in \eqref{eq:general-Hamiltonian-system} without changing the
density evolution or the velocity field. 
This gauge need not have zero spatial mean; identities involving the
normalized pseudoinverse are understood modulo spatial constants.

\paragraph{Legendre duality and the mechanical Lagrangian.}

For fixed $\rho$, $\mathcal H_0(\rho,\cdot)$ is strictly convex on
potential classes by Remark~\ref{rem:kinetic-energy-convexity}.
For $\sigma=-L_\rho\phi$, \eqref{eq:Hamiltonian-Legendre-relation}
and \eqref{eq:Hamiltonian-variational-derivatives} identify
$\phi=(-L_\rho)^\dagger\sigma$ as the unique maximizing potential
modulo constants. Evaluating the transform at this potential gives
\begin{equation*}
\mathcal L_0(\rho,\sigma)=\tfrac12\mathcal A_\rho(\phi,\phi)-\mathcal F(\rho)
=\tfrac12G_\rho(\sigma,\sigma)-\mathcal F(\rho).
\end{equation*}
Equivalently, the kinetic energies are Legendre dual:
\begin{equation*}
\mathcal K(\rho,\phi)
=\sup_{\tau\in T_\rho^{\mathrm{dens}}\mathcal P_+(M)}
\left\{\langle\phi,\tau\rangle_\rho-\tfrac12G_\rho(\tau,\tau)\right\}.
\end{equation*}
The kinetic-energy Legendre transform thus agrees with the musical maps
\eqref{eq:metric-Legendre-maps} in its identification of the variables:
$\sigma=\phi^\sharp$ and $\phi=\sigma^\flat$.

Writing $f_\rho=\delta\mathcal F/\delta\rho(\rho)$, the remaining
identities in \eqref{eq:Hamiltonian-Legendre-relation} give
\[
\begin{aligned}
D_\sigma\mathcal L_0(\rho,\sigma)[\tau]
&=\langle\phi,\tau\rangle_\rho=G_\rho(\sigma,\tau),\\
D_\rho\mathcal L_0(\rho,\sigma)[\eta]
&=-\int_M
\bigl[\Lambda_\rho(\phi,\phi)+f_\rho\bigr]\eta\,\mathrm dx,
\end{aligned}
\]
where the density derivative is taken at fixed $\sigma$.

\paragraph{Euler--Lagrange and nonlocal Newton formulations.}

Along a density curve, take
$\partial_t\rho_t=-L_{\rho_t}\phi_t$. Evaluating the derivatives at
$(\rho_t,\partial_t\rho_t)$ gives
\begin{equation*}
\begin{aligned}
\frac{\delta\mathcal L_0}{\delta\sigma}(\rho_t,\partial_t\rho_t)
&=\phi_t=(-L_{\rho_t})^\dagger\partial_t\rho_t,\\
\frac{\delta\mathcal L_0}{\delta\rho}(\rho_t,\partial_t\rho_t)
&=-\Lambda_{\rho_t}(\phi_t,\phi_t)-f_{\rho_t}.
\end{aligned}
\end{equation*}
Substitution into \eqref{eq:general-Euler-Lagrange} recovers the
nonlocal Hamilton--Jacobi equation in \eqref{eq:general-Hamiltonian-system}.
Together with the velocity relation, this identifies the mechanical
Hamiltonian and Lagrangian dynamics. Their covariant forms are as follows.

\begin{proposition}[Nonlocal Newton formulation]
\label{prop:HJ-Newton-equivalence}
Let $\rho_t$ be a smooth density curve with
$\partial_t\rho_t=-L_{\rho_t}\phi_t$, and assume that $J_{\rho_t}$ is
irreducible. Then \eqref{eq:general-Hamiltonian-system} is equivalent
to either of the nonlocal Newton equations
\begin{align}
\frac{D}{\mathrm dt}\partial_t\rho_t
&=L_{\rho_t}f_{\rho_t},
\label{eq:Newton-covariant}\\
D_t\Nabla\phi_t&=-\Nabla f_{\rho_t}.
\label{eq:Newton-potential}
\end{align}
\end{proposition}

\begin{proof}
In Corollary~\ref{prop:density-covariant}, take
$\tau_t=\partial_t\rho_t=-L_{\rho_t}\phi_t$ and $\psi_t=\phi_t$.
The two operator-variation terms in \eqref{eq:density-covariant}
then cancel, giving
\begin{equation}
\label{eq:covariant-acceleration-density}
\frac{D}{\mathrm dt}\partial_t\rho_t
=-L_{\rho_t}
\bigl(\partial_t\phi_t+\Lambda_{\rho_t}(\phi_t,\phi_t)\bigr).
\end{equation}
Using \eqref{eq:Riemannian-gradient}, we obtain
\[
\begin{aligned}
\frac{D}{\mathrm dt}\partial_t\rho_t-L_{\rho_t}f_{\rho_t}
&=-L_{\rho_t}
\bigl(\partial_t\phi_t+\Lambda_{\rho_t}(\phi_t,\phi_t)\bigr)
-L_{\rho_t}f_{\rho_t}\\
&=-L_{\rho_t}
\bigl(\partial_t\phi_t+\Lambda_{\rho_t}(\phi_t,\phi_t)
+f_{\rho_t}\bigr).
\end{aligned}
\]
Irreducibility identifies $\ker L_{\rho_t}$ with the spatial constants.
Hence \eqref{eq:Newton-covariant} is equivalent to the nonlocal Hamilton--Jacobi
equation in \eqref{eq:general-Hamiltonian-system}; the continuity equation
already holds by the velocity relation.

For the velocity-field representation, the transported connection
\eqref{eq:pushforward-covariant} and the gradient correspondence give
\[
\frac{D}{\mathrm dt}\partial_t\rho_t
=\mathbf I_{\rho_t}(D_t\Nabla\phi_t),
\qquad
L_{\rho_t}f_{\rho_t}
=-\mathbf I_{\rho_t}(\Nabla f_{\rho_t}).
\]
In particular, applying $\mathbf I_{\rho_t}^{-1}$ to
\eqref{eq:covariant-acceleration-density} gives
\begin{equation}
\label{eq:covariant-acceleration-potential}
D_t\Nabla\phi_t
=\Nabla\bigl(\partial_t\phi_t+\Lambda_{\rho_t}(\phi_t,\phi_t)\bigr).
\end{equation}
The same isometry identifies \eqref{eq:Newton-covariant} with
\eqref{eq:Newton-potential}, proving the second equivalence.
\end{proof}

\paragraph{Second-order density equation.}

By \eqref{eq:covariant-definition-density},
\begin{equation*}
\frac{D}{\mathrm dt}\partial_t\rho_t
=\partial_{tt}\rho_t
+\Gamma_W(\partial_t\rho_t,\partial_t\rho_t).
\end{equation*}
Combining this identity with \eqref{eq:Newton-covariant} gives
\begin{equation*}
\partial_{tt}\rho_t
+\Gamma_W(\partial_t\rho_t,\partial_t\rho_t)
=L_{\rho_t}f_{\rho_t}.
\end{equation*}
For $\sigma=-L_\rho\phi$, setting $\tau=\sigma$ and $\psi=\phi$
in \eqref{eq:Christoffel-density} gives, with
$\phi=-L_\rho^\dagger\sigma$,
\begin{equation}
\label{eq:Christoffel-diagonal}
\Gamma_W(\sigma,\sigma)=(DL_\rho[\sigma])\phi-L_\rho\Lambda_\rho(\phi,\phi)=-(DL_\rho[\sigma])L_\rho^\dagger\sigma
-L_\rho\Lambda_\rho
(L_\rho^\dagger\sigma,L_\rho^\dagger\sigma).
\end{equation}
Substituting \eqref{eq:Christoffel-diagonal}
with $\sigma=\partial_t\rho_t$ gives
\begin{equation*}
\partial_{tt}\rho_t
-(DL_{\rho_t}[\partial_t\rho_t])
L_{\rho_t}^\dagger\partial_t\rho_t -L_{\rho_t}\Lambda_{\rho_t}
\bigl(L_{\rho_t}^\dagger\partial_t\rho_t,
L_{\rho_t}^\dagger\partial_t\rho_t\bigr)
=L_{\rho_t}f_{\rho_t}.
\end{equation*}
This is the second-order Euler--Lagrange equation for $\mathcal L_0$.
Here the operator derivative acts at a fixed potential; in particular,
it does not differentiate $L_{\rho_t}^\dagger\partial_t\rho_t$.
As with the preceding pseudoinverse formulas, this expression is
used on the range of the operator.

For $\mathcal F\equiv0$, the two nonlocal Newton formulations reduce to the
geodesic equations
\[
\frac{D}{\mathrm dt}\partial_t\rho_t=0,
\qquad
D_t\Nabla\phi_t=0.
\]

\subsubsection{Gradient flows as Hamiltonian systems}
\label{subsubsec:gradient-Hamiltonian-lift}

Let $\mathcal F$ be a twice continuously differentiable density
functional, whose first variation
$f_\rho=\frac{\delta\mathcal F}{\delta\rho}(\rho)$ has a smooth
representative depending smoothly on $\rho$ in $C^\infty(M)$.
Define its squared gradient norm by
\begin{equation}
\label{eq:nl-Fisher}
\begin{aligned}
\mathscr F_{\mathcal F}(\rho)
&=G_\rho\bigl(\mathrm{grad}^{\mathrm{dens}}_\rho\mathcal F,
\mathrm{grad}^{\mathrm{dens}}_\rho\mathcal F\bigr)\\
&=\mathcal A_\rho(f_\rho,f_\rho).
\end{aligned}
\end{equation}
Consider the mechanical Hamiltonian
\begin{equation}\label{Hamiltonian-grad}
\mathcal H_{\mathcal F}(\rho,\phi)
:=\mathcal K(\rho,\phi)-\tfrac12\mathscr F_{\mathcal F}(\rho)
=\tfrac12\mathcal A_\rho(\phi,\phi)
-\tfrac12\mathscr F_{\mathcal F}(\rho).
\end{equation}

\begin{proposition}[Hamiltonian lift of a gradient flow]
\label{prop:gradient-Hamiltonian-lift}
Assume $f_{\rho}\in\mathscr D(L_\rho)$. Assume also that
$\delta\mathscr F_{\mathcal F}/\delta\rho$ has a smooth
representative depending smoothly on $\rho$. Every smooth solution
of the negative gradient flow
\begin{equation}
\label{eq:scaled-gradient-flow}
\partial_t\rho_t
=-\mathrm{grad}^{\mathrm{dens}}_{\rho_t}\mathcal F
=L_{\rho_t}f_{\rho_t}
\end{equation}
satisfies the nonlocal Newton equation
\begin{equation}
\label{eq:gradient-flow-Newton}
\frac{D}{\mathrm dt}\partial_t\rho_t
=\frac12\mathrm{grad}^{\mathrm{dens}}_{\rho_t}
\mathscr F_{\mathcal F}
=-\frac12L_{\rho_t}
\frac{\delta\mathscr F_{\mathcal F}}{\delta\rho}(\rho_t).
\end{equation}
It admits a momentum potential $\phi_t$ for which it is a trajectory
of $\mathcal H_{\mathcal F}$:
\begin{equation}
\label{eq:nl-Hamilton-system}
\begin{cases}
\displaystyle
\partial_t\rho_t=-L_{\rho_t}\phi_t,\\[1mm]
\displaystyle
\partial_t\phi_t+\Lambda_{\rho_t}(\phi_t,\phi_t)
=\frac12
\frac{\delta\mathscr F_{\mathcal F}}{\delta\rho}(\rho_t).
\end{cases}
\end{equation}
The potential equation is understood modulo spatial constants.
\end{proposition}

\begin{proof}
For fixed density tangent vectors $\sigma,\tau$, metric compatibility
and the defining property of the gradient give
\[
G_\rho\bigl(\nabla_\sigma\mathrm{grad}^{\mathrm{dens}}\mathcal F,\tau\bigr)
=D^2\mathcal F(\rho)[\sigma,\tau]
-D\mathcal F(\rho)[\Gamma_W(\sigma,\tau)].
\]
The right-hand side is symmetric in $\sigma,\tau$. Differentiating
\eqref{eq:nl-Fisher} and using this symmetry therefore yields
\[
D\mathscr F_{\mathcal F}(\rho)[\sigma]
=2G_\rho\bigl(\nabla_\sigma\mathrm{grad}^{\mathrm{dens}}\mathcal F,
\mathrm{grad}^{\mathrm{dens}}\mathcal F\bigr)=2G_\rho\bigl(\nabla_{\mathrm{grad}^{\mathrm{dens}}\mathcal F}
\mathrm{grad}^{\mathrm{dens}}\mathcal F,\sigma\bigr).
\]
Thus
$\nabla_{\mathrm{grad}^{\mathrm{dens}}\mathcal F}
\mathrm{grad}^{\mathrm{dens}}\mathcal F
=\frac12\mathrm{grad}^{\mathrm{dens}}\mathscr F_{\mathcal F}$.
Taking a covariant time derivative of \eqref{eq:scaled-gradient-flow} gives
\[
\frac{D}{\mathrm dt}\partial_t\rho_t
=-\nabla_{\partial_t\rho_t}
\mathrm{grad}^{\mathrm{dens}}\mathcal F
=\frac12\mathrm{grad}^{\mathrm{dens}}
\mathscr F_{\mathcal F}.
\]
The gradient formula \eqref{eq:Riemannian-gradient} gives the operator
expression in \eqref{eq:gradient-flow-Newton}.
Choose the momentum potential
\begin{equation}
\label{eq:gradient-flow-momentum-graph}
\phi_t=-f_{\rho_t}.
\end{equation}
Then \eqref{eq:scaled-gradient-flow} gives
$\partial_t\rho_t=-L_{\rho_t}\phi_t$.
Applying Proposition~\ref{prop:HJ-Newton-equivalence} with potential
energy $-\frac12\mathscr F_{\mathcal F}$ yields the potential equation
in \eqref{eq:nl-Hamilton-system}, modulo spatial constants.
\end{proof}

The momentum constraint \eqref{eq:gradient-flow-momentum-graph} selects
a zero-Hamiltonian \eqref{Hamiltonian-grad}; arbitrary Hamiltonian trajectories need
not satisfy the original first-order gradient equation.

\section{Examples}
\label{sec:examples}

We organize the examples into gradient flows and Hamiltonian flows,
using the density energies below. We also compare density dynamics
with classical phase-space Vlasov dynamics, which uses a different
underlying geometry.

Euclidean examples below are explicit extensions of the closed-manifold
setting, with sufficient decay assumed on noncompact spaces.
Calculations are performed on smooth fields in the relevant domains; singular
integrals are interpreted in the principal-value or Dirichlet-form sense.

\paragraph{Density energies.}

Let $V\in C^\infty(M)$ be a real external potential and let
$W\in C^\infty(M\times M)$ be a symmetric interaction kernel,
$W(x,y)=W(y,x)$, distinct from the jump kernel $K$. We use
\begin{align}
\mathcal E(\rho)
&=\int_M\rho(x)\log\rho(x)\,\mathrm dx,
\label{eq:Boltzmann-entropy}\\
\mathcal R_m(\rho)
&=\frac1{m-1}\int_M\bigl(\rho(x)^m-\rho(x)\bigr)\,\mathrm dx,
\qquad m>0,\quad m\ne1,
\label{eq:Renyi-entropy}\\
\mathcal V(\rho)
&=\int_MV(x)\rho(x)\,\mathrm dx,
\label{eq:examples-Vlasov-density-energy}\\
\mathcal W(\rho)
&=\frac12\iint_{M\times M}
W(x,y)\rho(x)\rho(y)\,\mathrm dx\,\mathrm dy.
\label{eq:examples-interaction-energy}
\end{align}
Here $\mathcal R_m$ uses the power-integral convention for
R\'enyi entropy common in transportation theory
\cite[Remark~1.3]{Erbar+Maas_2014_Porous}; the mass-linear term
normalizes it so that $\mathcal R_m\to\mathcal E$ as $m\to1$.
For a smooth function $h$, we write
\[
(W*h)(x):=\int_MW(x,y)h(y)\,\mathrm dy;
\]
this denotes the kernel action and does not require translation invariance.

\subsection{Gradient-flow examples}
\label{subsec:gradient-flow-examples}

{
We apply the gradient-flow equation \eqref{eq:density-gradient-flow}
and the general dissipation identity \eqref{eq:examples-entropy-dissipation}.
Variational derivatives with respect to a probability density are
understood modulo spatial constants. For an internal energy
$\mathcal F_f(\rho)=\int_Mf(\rho)\,\mathrm dx$, this yields
\begin{equation*}
\partial_t\rho(x)=2\mathrm{p.v.}\int_M[f'(\rho(y))-f'(\rho(x))]\theta(\rho(x),\rho(y))K(x,y)\,\mathrm dy.
\end{equation*}
The mobility is kept fixed unless a change is explicitly specified.
}

\subsubsection{Boltzmann entropy}
\label{subsec:logarithmic-mean-fractional-heat-revised}

The entropy \eqref{eq:Boltzmann-entropy} has first variation
\begin{equation}
\label{eq:examples-Boltzmann-first-variation}
\frac{\delta\mathcal E}{\delta\rho}(\rho)=\log\rho+1.
\end{equation}
Assume that $\log\rho\in\mathscr H_\rho$. Since $L_\rho 1=0$,
the negative gradient flow for the nonlocal metric is
\begin{equation}
\label{eq:entropy-gradient-flow}
\partial_t\rho_t(x)
=
2\mathrm{p.v.}\int_M
\bigl(
\log\rho_t(y)-\log\rho_t(x)
\bigr)
\theta_{\rho_t}(x,y)K(x,y)\,
\mathrm dy.
\end{equation}

\paragraph{Logarithmic-mean mobility and heat flow.}

For the logarithmic mean \eqref{eq:logarithmic-mean-definition},
the key identity is
\begin{equation}
\label{eq:examples-logarithmic-chain-rule}
\theta_{\log}(a,b)(\log b-\log a)=b-a.
\end{equation}

\begin{proposition}[Logarithmic-mean gradient flow and fractional diffusion]
\label{prop:logarithmic-mean-fractional-heat-revised}
The negative gradient flow of $\mathcal E$ for the logarithmic-mean
metric is the linear nonlocal heat equation
\begin{equation*}
\partial_t\rho_t(x)
=2\mathrm{p.v.}\int_M\bigl(\rho_t(y)-\rho_t(x)\bigr)K(x,y)\,\mathrm dy
=L_K\rho_t(x),
\end{equation*}
where $L_K$ denotes the generator with mobility $\theta\equiv1$.
\end{proposition}

\begin{proof}
By \eqref{eq:density-gradient-flow},
\eqref{eq:examples-Boltzmann-first-variation}, and $L_\rho1=0$,
\[
\begin{aligned}
\partial_t\rho_t(x)=L_{\rho_t}\log\rho_t(x)
&=2\mathrm{p.v.}\int_M
\theta_{\log}(\rho_t(x),\rho_t(y))
\bigl(\log\rho_t(y)-\log\rho_t(x)\bigr)K(x,y)\,\mathrm dy\\
&=2\mathrm{p.v.}\int_M\bigl(\rho_t(y)-\rho_t(x)\bigr)K(x,y)\,\mathrm dy,
\end{aligned}
\]
where the last equality follows from
\eqref{eq:examples-logarithmic-chain-rule}.
\end{proof}

\paragraph{Entropy production.}

By \eqref{eq:examples-entropy-dissipation} with $\mathcal F=\mathcal E$,
\[
\frac{\mathrm d}{\mathrm dt}\mathcal E(\rho_t)
=-\mathcal A_{\rho_t}(\log\rho_t,\log\rho_t)\leq0.
\]
On a closed $M$, the symmetric heat flow has the uniform equilibrium
$\rho_\infty=1/\operatorname{vol}(M)$. With the statistical-physics
convention $S(\rho)=-k_{\mathrm B}\mathcal E(\rho)$, up to an additive
constant, the same identity gives
\[
\frac{\mathrm d}{\mathrm dt}S(\rho_t)
=k_{\mathrm B}\mathcal A_{\rho_t}(\log\rho_t,\log\rho_t)\geq0.
\]
This is the nonlocal $H$-theorem: for diffusion with uniform equilibrium,
it expresses entropy production in the sense of the second law of
thermodynamics.

\paragraph{Fractional heat flow.}
On the unit flat torus, take the periodized kernel
\begin{equation}\label{eq:examples-fractional-kernel}
K_s(x,y)=c_{d,s}\sum_{n\in\mathbb Z^d}
\frac{1}{|x-y+n|^{d+2s}},\qquad 0<s<1,
\end{equation}
for $x\ne y$, where
\[
c_{d,s}=\frac{4^s\Gamma(d/2+s)}{\pi^{d/2}|\Gamma(-s)|}.
\]
The sum is independent of the chosen lifts of $x$ and $y$ to
$\mathbb R^d$. The pointwise formula for the torus fractional
Laplacian, after rescaling the period to one, is
\cite[Theorem~1.5]{Roncal+Stinga_2016_Fractional}
\[
(-\Delta_{\mathbb T^d})^sf(x)
=\operatorname{p.v.}\int_{\mathbb T^d}
[f(x)-f(y)]K_s(x,y)\,\mathrm dy.
\]
Its Fourier multiplier on $e^{2\pi i n\cdot x}$ is
$(2\pi|n|)^{2s}$. With the generator normalization used here,
\[
L_{K_s}f=-2(-\Delta_{\mathbb T^d})^sf,
\qquad
\partial_t\rho_t+2(-\Delta_{\mathbb T^d})^s\rho_t=0.
\]
The second identity is the logarithmic-mean entropy-flow equation.
For the related gradient-flow interpretation of homogeneous jump
semigroups on $\mathbb R^d$ under the assumptions used there, see
\cite[Theorem~1.4 and Corollary~1.5]{Erbar_2014_Gradient}.

\paragraph{Geometric-mean mobility and linearization.}

Take $\theta$ as the geometric mean in \eqref{eq:geometric-mean-definition}.
Then \eqref{eq:entropy-gradient-flow} becomes
\begin{equation}
\label{eq:geometric-mean-entropy-flow}
\partial_t\rho_t(x)
=
2\mathrm{p.v.}\int_M
\sqrt{\rho_t(x)\rho_t(y)}
\log\left(
\frac{\rho_t(y)}{\rho_t(x)}
\right)
K(x,y)\,
\mathrm dy.
\end{equation}
Unlike the logarithmic-mean case, the density dependence does not
cancel, so \eqref{eq:geometric-mean-entropy-flow} is genuinely
nonlinear.
Its entropy dissipation is a special case of
\eqref{eq:examples-entropy-dissipation}.

We examine the behavior near the constant equilibrium.
Assume that $M=\mathbb T^d$ is equipped with normalized Lebesgue
measure and write
\begin{equation*}
\rho_t^\epsilon
=
1+\epsilon u_t^\epsilon,
\qquad
\int_Mu_t^\epsilon\,\mathrm dx=0,
\end{equation*}
with $1+\epsilon u_t^\epsilon>0$.
Substituting directly into
\eqref{eq:geometric-mean-entropy-flow} and using the fact that
for fixed $a,b\in\mathbb R$,
\begin{equation*}
\lim_{\epsilon\to0}
\frac{
\sqrt{(1+\epsilon a)(1+\epsilon b)}
}{\epsilon}
\log\left(
\frac{1+\epsilon b}{1+\epsilon a}
\right)
=
b-a.
\end{equation*}
we see that $u^\epsilon\to u$ in a topology that allows passage to
the limit in the nonlocal integral. Thus, the linearized equation is
\begin{equation*}
\partial_tu_t(x)
=
2\mathrm{p.v.}\int_M
\bigl(
u_t(y)-u_t(x)
\bigr)
K(x,y)\,
\mathrm dy.
\end{equation*}
For the periodized fractional kernel corresponding to \eqref{eq:examples-fractional-kernel},
the linearized equation becomes
\begin{equation*}
\partial_tu_t
=
-2(-\Delta_{\mathbb T^d})^s u_t.
\end{equation*}

Thus, for the same entropy $\mathcal E$, the logarithmic mean gives a
linear nonlocal heat equation, while the geometric mean produces a
nonlinear entropy-dissipating flow with the same linearization at
constant density.

\subsubsection{R\'enyi entropy}

For the R\'enyi entropy \eqref{eq:Renyi-entropy},
\[
\frac{\delta\mathcal R_m}{\delta\rho}=\frac{m\rho^{m-1}-1}{m-1},\quad \partial_t\rho=\frac{m}{m-1}L_\rho(\rho^{m-1}).
\]
This is the R\'enyi-entropy flow for the previously fixed mobility.
To obtain a nonlocal porous-medium or fast-diffusion equation,
choose instead
\begin{equation*}
\theta_m(a,b)
=
\begin{cases}
\displaystyle\frac{m-1}{m}
\frac{b^m-a^m}{b^{m-1}-a^{m-1}},&a\ne b,\\[2mm]
a,&a=b.
\end{cases}
\end{equation*}
The continuous extension at the diagonal is understood.
Since
\[
\theta_m(a,b)\frac{m}{m-1}
\bigl(b^{m-1}-a^{m-1}\bigr)=b^m-a^m,
\]
the flow reduces to
\begin{equation}
\label{eq:nl-porous-medium}
\partial_t\rho=L_K(\rho^m).
\end{equation}
For the fractional kernel $K=K_s$, this reads
\[
\partial_t\rho+2(-\Delta)^s(\rho^m)=0.
\]
The cases $m>1$ and $0<m<1$ are the porous-medium and fast-diffusion
regimes, respectively.
Moreover,
\[
\lim_{m\to1}\mathcal R_m(\rho)=\mathcal E(\rho),
\qquad
\lim_{m\to1}\theta_m(a,b)=\theta_{\log}(a,b).
\]
The power-mobility construction is already present in the
finite-state setting of
\cite[Theorem~1.1 and Section~3.3]{Erbar+Maas_2014_Porous}.
The continuum identities above are direct calculations under our
smoothness assumptions, not an invocation of a finite-state
existence theorem.

\begin{remark}[Dependence of the geometry on the mobility]
\label{rem:nl-mobility-scope}
Changing $\theta$ changes $L_\rho$, $\mathcal A_\rho$, $G_\rho$, and
$\Lambda_\rho$ simultaneously.
Thus \eqref{eq:nl-porous-medium} belongs to a family of
entropy--metric pairs, rather than to a single fixed metric with
varying entropy.
The mean $\theta_m$ is concave for $0<m\leq2$; see
\cite[Section~3.3]{Erbar+Maas_2014_Porous}.
For $m>2$, the direct expansion
\[
\theta_m(1-h,1+h)
=1+\frac{m-2}{3}h^2+O(h^4)
\qquad(h\to0)
\]
shows that it is not concave.
Although the displayed positive-density identities still hold for
$m>2$, results requiring a concave mobility cannot be applied in
that range.
If a metric construction additionally requires
$\theta(0,b)=0$, this also needs checking:
\[
\theta_m(0,b)=\frac{m-1}{m}b>0
\qquad(m>1,\ b>0).
\]
No conclusion about extension to vanishing densities or metric
completion is implicit in these examples.
\end{remark}

\subsubsection{Nonlocal mean-field equations}

\label{ex:nl-interaction-gradient}
Using \eqref{eq:Boltzmann-entropy},
\eqref{eq:examples-Vlasov-density-energy}, and
\eqref{eq:examples-interaction-energy}, consider
\[
\mathcal F(\rho)=\mathcal E(\rho)+\mathcal V(\rho)+\mathcal W(\rho).
\]
By symmetry of $W$, a representative of its first variation is
\[
\frac{\delta\mathcal F}{\delta\rho}(\rho)=\log\rho+1+V+W*\rho.
\]
Consequently,
\begin{equation*}
\partial_t\rho=L_\rho\bigl(\log\rho+V+W*\rho\bigr).
\end{equation*}
With the logarithmic mean, this becomes
\[
\partial_t\rho=L_K\rho+L_\rho(V+W*\rho).
\]
This is the nonlocal mean-field drift--diffusion equation generated
by the specified energy and mobility.

\paragraph{Relative entropy as a special case.}

For $W=0$, define the Gibbs density
\[
\pi=Z^{-1}e^{-V},
\qquad Z=\int_Me^{-V}\,\mathrm dx.
\]
The relative entropy is
\[
\mathcal E(\rho\mid\pi)
=\int_M\rho\log\frac{\rho}{\pi}\,\mathrm dx
=\mathcal E(\rho)+\mathcal V(\rho)+\log Z.
\]
It differs from $\mathcal F$ with $W=0$ only by a constant,
so both generate the flow
$\partial_t\rho=L_\rho(\log\rho+V)$. For the logarithmic mean,
\begin{equation*}
\partial_t\rho=L_K\rho+L_\rho V.
\end{equation*}
The density $\pi$ is stationary because $\log\pi+V$ is constant.
This is a nonlocal drift--diffusion analogue of the Fokker--Planck
equation. Since $L_\rho V$ generally depends nonlinearly on $\rho$,
it is not in general a linear jump-process forward equation.
For a fractional kernel, the drift is still $L_\rho V$, not
$\nabla\cdot(\rho\nabla V)$.

\subsubsection{Nonlocal Keller--Segel flow}

\label{ex:ks-nonlocal-mixed-gradient}
Let $M=\mathbb T^d$ carry normalized Lebesgue measure $\mathrm dx$.
Fix $d_1,d_2,\chi,\beta>0$ and $\alpha\geq0$.
We consider smooth $u>0$ with $\int_Mu\,\mathrm dx=1$ and a smooth
signal concentration $v$, with no mass constraint on $v$.
Both components use the same symmetric jump kernel $K$.
The density-dependent operator $L_u$ means $L_\rho$ evaluated at
$\rho=u$, whereas $L_K$ has mobility identically equal to one.

Consider the free energy
\begin{equation}
\label{eq:ks-free-energy}
\begin{aligned}
\mathcal F_{\mathrm{KS}}(u,v)
={}&\frac{d_1}{\chi}\mathcal E(u)
+\int_M\left[-uv+\frac{\alpha}{2\beta}v^2\right]\,\mathrm dx\\
&+\frac{d_2}{2\beta}\iint_{M\times M}
\bigl(v(y)-v(x)\bigr)^2K(x,y)\,\mathrm dx\,\mathrm dy.
\end{aligned}
\end{equation}
Here the entropy is given by \eqref{eq:Boltzmann-entropy}, and the
signal energy is a jump Dirichlet energy.
The normalization of $L_K$ gives
\[
\iint_{M\times M}\bigl(v(y)-v(x)\bigr)^2K(x,y)\,\mathrm dx\,\mathrm dy
=-\int_MvL_Kv\,\mathrm dx.
\]
By symmetry of $K$, the first variations are
\begin{equation*}
\begin{aligned}
\frac{\delta\mathcal F_{\mathrm{KS}}}{\delta u}
&=\frac{d_1}{\chi}(\log u+1)-v,\\
\frac{\delta\mathcal F_{\mathrm{KS}}}{\delta v}
&=-u-\frac{d_2}{\beta}L_Kv+\frac{\alpha}{\beta}v.
\end{aligned}
\end{equation*}
Only the first variation with respect to $u$ is understood modulo
spatial constants. 

On $\mathcal P_+(M)\times C^\infty(M)$ use the product metric
\begin{equation}
\label{eq:ks-nonlocal-product-metric}
\tfrac1\chi G\ \oplus\ \tfrac1\beta
\langle\cdot,\cdot\rangle_{L^2(M)}.
\end{equation}
Then the negative gradient flow of \eqref{eq:ks-free-energy} for
\eqref{eq:ks-nonlocal-product-metric} is
\begin{equation*}
\begin{cases}
\displaystyle
\partial_tu=\chi L_u\frac{\delta\mathcal F_{\mathrm{KS}}}{\delta u}
=d_1L_u\log u-\chi L_uv,\\[2mm]
\displaystyle
\partial_tv=-\beta\frac{\delta\mathcal F_{\mathrm{KS}}}{\delta v}
=d_2L_Kv-\alpha v+\beta u.
\end{cases}
\end{equation*}
For the logarithmic-mean mobility, $L_u\log u=L_Ku$, and hence
\begin{equation*}
\begin{cases}
\partial_tu=d_1L_Ku-\chi L_uv,\\[1mm]
\partial_tv=d_2L_Kv-\alpha v+\beta u.
\end{cases}
\end{equation*}
Symmetry gives $\int_M\partial_tu\,\mathrm dx=0$, whereas
\[
\frac{\mathrm d}{\mathrm dt}\int_Mv\,\mathrm dx
=-\alpha\int_Mv\,\mathrm dx+\beta.
\]
This explains the use of an $L^2$ factor, rather than a second
fixed-mass density space, for the signal.

The entropy coefficient and product weights in this example follow
the normalization of the hybrid formulation in
\cite[Section~1.2, p.~1555, equations~(1.3)--(1.4)]
{Blanchet+Carrillo+Kinderlehrer+Kowalczyk+Laurencot+Lisini_2015_Hybrid}.
That reference treats a local model; the jump signal energy and the
fully nonlocal equations here are a separate construction, verified
by the above first-variation calculation. These identities concern
smooth solutions and do not assert a nonlocal existence theorem,
a critical-mass threshold, or convergence of a variational scheme.

\subsection{Hamiltonian-flow examples}
\label{subsec:Hamiltonian-flow-examples}

We now turn to canonical density--potential dynamics, beginning
with interaction energy and the quadratic kinetic energy of the
nonlocal metric. The Hopf--Cole and Madelung transformations then
lead to hyperbolic and trigonometric kinetic energies. Their
conjugate potentials need not coincide with the metric velocity
potentials representing the density curves.

\subsubsection{Nonlocal Vlasov equation}
\label{subsec:nonlocal-Vlasov-revised}

The nonlocal Hamilton--Jacobi and nonlocal Newton equations
accommodate the potential energy $\mathcal V+\mathcal W$ from
\eqref{eq:examples-Vlasov-density-energy} and
\eqref{eq:examples-interaction-energy}.

\begin{proposition}[Density dynamics with interaction energy]
\label{prop:nonlocal-Vlasov-revised}
For any positive symmetric mobility, the density Hamiltonian
$\mathcal K+\mathcal V+\mathcal W$ generates
\begin{equation}
\label{eq:examples-Vlasov-density-system}
\begin{cases}
\displaystyle
\partial_t\rho_t=(-L_{\rho_t})\phi_t,
\\[0.5em]
\displaystyle
\partial_t\phi_t
+
\Lambda_{\rho_t}(\phi_t,\phi_t)
+
V
+
\displaystyle\int_MW(\,\cdot\,,y)\rho_t(y)\,\mathrm dy
=C(t).
\end{cases}
\end{equation}
Equivalently, the corresponding smooth density curve satisfies the nonlocal Newton
equation
\begin{equation}
\label{eq:examples-Vlasov-density-Newton}
\frac{D}{\mathrm dt}\partial_t\rho_t
=
L_{\rho_t}
\left(
V+\int_MW(\,\cdot\,,y)\rho_t(y)\,\mathrm dy
\right).
\end{equation}
\end{proposition}

\begin{proof}
For a density perturbation $\sigma$, symmetry of $W$ gives
\begin{align*}
D(\mathcal V+\mathcal W)(\rho)[\sigma]
&=
\int_MV(x)\sigma(x)\,\mathrm dx
+
\frac12\iint W(x,y)
\bigl[\sigma(x)\rho(y)+\rho(x)\sigma(y)\bigr]\,\mathrm dx\,\mathrm dy
\\
&=
\int_M
\left[
V(x)+\int_MW(x,y)\rho(y)\,\mathrm dy
\right]\sigma(x)\,\mathrm dx.
\end{align*}
Hence
\begin{equation*}
\frac{\delta(\mathcal V+\mathcal W)}{\delta\rho}(\rho)(x)
=V(x)+\int_MW(x,y)\rho(y)\,\mathrm dy.
\end{equation*}
The density Hamilton equations and
\eqref{eq:examples-Vlasov-density-Newton} follow by applying
Theorem~\ref{thm:quadratic-Hamiltonian-equations} and
Proposition~\ref{prop:HJ-Newton-equivalence} to $\mathcal F=\mathcal V+\mathcal W$.
\end{proof}

\paragraph{Phase-space interpretation.}

On the other hand, when $M=\mathbb R^d$, the
classical phase-space Vlasov equation for
$f_t(x,v)$ is
\begin{equation}
\label{eq:examples-Vlasov-equation}
\partial_t f_t
+
v\cdot\nabla_x f_t
-
\nabla_x\bigl(V+W*\rho_{f_t}\bigr)\cdot\nabla_v f_t
=0,
\qquad
\rho_{f_t}(x)=\int_{\mathbb R^d}f_t(x,v)\,\mathrm dv.
\end{equation}
It is generated by the phase-space Hamiltonian
\begin{equation*}
\mathscr H(f)
=\iint_{\mathbb R^{2d}}\frac{|v|^2}{2}f(x,v)\,\mathrm dx\,\mathrm dv
+\mathcal V(\rho_f)+\mathcal W(\rho_f),
\end{equation*}
whose first variation is
\begin{equation}
\label{eq:examples-Vlasov-phase-first-variation}
\frac{\delta\mathscr H}{\delta f}(x,v)
=
\frac{|v|^2}{2}+V(x)+(W*\rho_f)(x).
\end{equation}
With the canonical bracket
$
\{a,b\}_{x,v}=\nabla_xa\cdot\nabla_vb-\nabla_va\cdot\nabla_xb
$, equation \eqref{eq:examples-Vlasov-equation} is equivalent to
\begin{equation}
\label{eq:examples-Vlasov-Poisson}
\partial_t f
+
\left\{f,\frac{\delta\mathscr H}{\delta f}\right\}_{x,v}
=0.
\end{equation}
For the phase-space functional, variation with respect to $f$ gives
\eqref{eq:examples-Vlasov-phase-first-variation}.  Since
$
\nabla_v(\delta\mathscr H/\delta f)=v
$
and
$
\nabla_x(\delta\mathscr H/\delta f)
=\nabla_x(V+W*\rho_f)
$, expanding the Poisson bracket in
\eqref{eq:examples-Vlasov-Poisson} gives exactly
\eqref{eq:examples-Vlasov-equation}.

The relation between \eqref{eq:examples-Vlasov-density-system} and
\eqref{eq:examples-Vlasov-equation} is therefore structural rather than an
identification of the two geometries.  The first is a Hamiltonian flow on the
scalar density manifold constructed in this paper; the second is a
phase-space Lie--Poisson equation.  They share the same potential energy $\mathcal V+\mathcal W$.

\paragraph{Nonlocal Euler formulation.}

For a smooth tangent velocity
$v_t\in T_{\rho_t}^{\mathrm{vel}}\mathcal P_+(M)$, the nonlocal
Euler system associated with the potential energy $\mathcal V+\mathcal W$ is
\begin{equation}
\label{eq:examples-pressureless-Euler-Vlasov}
\begin{cases}
\displaystyle
\partial_t\rho_t+\Div(J_{\rho_t}v_t)=0,\\[0.6em]
\begin{aligned}
\partial_tv_t
&+\Nabla\left[
\int_M v_t(\,\cdot\,,y)^2
\partial_a\theta\bigl(\rho_t(\,\cdot\,),\rho_t(y)\bigr)
K(\,\cdot\,,y)\,\mathrm dy
\right]\\
&=-\Nabla\left[
V+\int_MW(\,\cdot\,,y)\rho_t(y)\,\mathrm dy
\right].
\end{aligned}
\end{cases}
\end{equation}
The first equation gives $\partial_t\rho_t=\mathbf I_{\rho_t}v_t$.
Taking the differentiated field to be the curve velocity in
\eqref{eq:general-covariant-potential} cancels its two
operator-variation terms. The second equation therefore writes out
$D_tv_t$, with the nonlinear term determined by the density-dependent mobility.

When $v_t=\Nabla\phi_t$, the first equation is
$\partial_t\rho_t=-L_{\rho_t}\phi_t$.
By Definition~\ref{def:density-variation-operator}, the integral in the
second equation is the density-variation term
$\Lambda_{\rho_t}(\phi_t,\phi_t)$, and
\eqref{eq:covariant-acceleration-potential} gives
\[
D_tv_t
=\partial_tv_t+\Nabla\Lambda_{\rho_t}(\phi_t,\phi_t)
=\Nabla\bigl(\partial_t\phi_t
+\Lambda_{\rho_t}(\phi_t,\phi_t)\bigr).
\]
Consequently, the velocity equation is equivalent, modulo spatial
constants, to the nonlocal Hamilton--Jacobi equation
\begin{equation*}
\partial_t\phi_t+\Lambda_{\rho_t}(\phi_t,\phi_t)
+V+\int_MW(\,\cdot\,,y)\rho_t(y)\,\mathrm dy=C(t).
\end{equation*}
Thus the potential representation of
\eqref{eq:examples-pressureless-Euler-Vlasov} coincides with
\eqref{eq:examples-Vlasov-density-system}.
This is an Euler formulation for the jump-kernel metric, not a
classical phase-space moment closure.

\subsubsection{Nonlocal Schr\"odinger bridge}
\label{subsec:fractional-Schrodinger-bridge}

\paragraph{Setup and Hopf--Cole transformation.}

For a symmetric reference kernel $K$, retain the normalization
\begin{equation}
\label{eq:examples-LcalK}
\frac12L_K f(x)=\mathrm{p.v.}\int_M\Nabla f(x,y)K(x,y)\,\mathrm dy.
\end{equation}
For the fractional kernel \eqref{eq:examples-fractional-kernel},
$\frac12L_{K_s}=-(-\Delta)^s$.
Use the fixed external potential $V$ from
\eqref{eq:examples-Vlasov-density-energy}.
The formal Schr\"odinger system consists of positive factors
$\Phi_t,\Phi_t^*$ satisfying
\begin{equation}
\label{eq:master}
\begin{cases}
\displaystyle
\partial_t\Phi_t=-\frac12L_K\Phi_t+V\Phi_t,\\[0.4em]
\displaystyle
-\partial_t\Phi_t^*=-\frac12L_K\Phi_t^*+V\Phi_t^*.
\end{cases}
\end{equation}
Given bridge endpoint densities $\rho_0,\rho_T$, the factors satisfy
$\Phi_0\Phi_0^*=\rho_0$ and $\Phi_T\Phi_T^*=\rho_T$.
Introduce the Hopf--Cole variables
\begin{equation}
\label{eq:colehopf}
\Phi_t=\sqrt{\rho_t}\,e^{\phi_t},
\qquad
\Phi_t^*=\sqrt{\rho_t}\,e^{-\phi_t},
\end{equation}
so that $\rho_t=\Phi_t\Phi_t^*$ and
$\phi_t=\frac12(\log\Phi_t-\log\Phi_t^*)$.
Throughout the bridge calculation, the mobility is the geometric mean:
$J_\rho(x,y)=\sqrt{\rho(x)\rho(y)}K(x,y)$.

\paragraph{Nonlocal continuity and potential equations.}

\begin{proposition}[Hopf--Cole representation]
\label{prop:bridge-Cole-Hopf}
Under \eqref{eq:colehopf}, the positive-factor system
\eqref{eq:master} is equivalent to
\begin{equation}
\label{eq:continuity}
\partial_t\rho_t+\Div\bigl(J_{\rho_t}\sinh(\Nabla\phi_t)\bigr)=0
\end{equation}
and
\begin{equation}
\label{eq:hj_raw}
\partial_t\phi_t(x)
=V(x)-\mathrm{p.v.}\int_M\left[
\sqrt{\frac{\rho_t(y)}{\rho_t(x)}}
\cosh\bigl(\Nabla\phi_t(x,y)\bigr)-1
\right]K(x,y)\,\mathrm dy.
\end{equation}
\end{proposition}

\begin{proof}
Differentiating the product and using \eqref{eq:master}, the
potential terms cancel:
\begin{align*}
\partial_t\rho_t(x)
&=\Phi_t^*(x)\partial_t\Phi_t(x)
+\Phi_t(x)\partial_t\Phi_t^*(x)\\
&=-\frac12\Phi_t^*(x)L_K\Phi_t(x)
+\frac12\Phi_t(x)L_K\Phi_t^*(x)\\
&=\mathrm{p.v.}\int_M\bigl[\Phi_t(x)\Phi_t^*(y)
-\Phi_t^*(x)\Phi_t(y)\bigr]K(x,y)\,\mathrm dy\\
&=-2\mathrm{p.v.}\int_M\sqrt{\rho_t(x)\rho_t(y)}
\sinh\bigl(\Nabla\phi_t(x,y)\bigr)K(x,y)\,\mathrm dy.
\end{align*}
Since $J_{\rho_t}\sinh(\Nabla\phi_t)$ is antisymmetric,
\eqref{eq:graph_div} gives \eqref{eq:continuity}.

For the logarithmic ratio,
\[
\partial_t\phi_t
=\frac12\left(
\frac{\partial_t\Phi_t}{\Phi_t}
-\frac{\partial_t\Phi_t^*}{\Phi_t^*}\right)=V-\frac14\left(
\frac{L_K\Phi_t}{\Phi_t}
+\frac{L_K\Phi_t^*}{\Phi_t^*}\right).
\]
Plugging in the definition of $L_K$ proves \eqref{eq:hj_raw}.
Conversely, differentiating the two factors in
\eqref{eq:colehopf} gives
\[
\frac{\partial_t\Phi_t}{\Phi_t}
=\frac{\partial_t\rho_t}{2\rho_t}+\partial_t\phi_t,
\qquad
\frac{\partial_t\Phi_t^*}{\Phi_t^*}
=\frac{\partial_t\rho_t}{2\rho_t}-\partial_t\phi_t.
\]
Substituting \eqref{eq:continuity} and \eqref{eq:hj_raw} recovers
\eqref{eq:master}.
\end{proof}

The amplitude term $\sqrt{\rho_t(y)/\rho_t(x)}-1$ contains a single increment of $\sqrt{\rho_t}$ and in general requires the principal value in \eqref{eq:hj_raw}. 
Separating $\cosh(\Nabla\phi)-1$ from the amplitude term gives
\begin{equation}
\label{eq:hj_split}
\partial_t\phi_t(x)
+\int_M\sqrt{\frac{\rho_t(y)}{\rho_t(x)}}
\bigl[\cosh\bigl(\Nabla\phi_t(x,y)\bigr)-1\bigr]K(x,y)\,\mathrm dy
=V(x)-\frac{L_K\sqrt{\rho_t}(x)}{2\sqrt{\rho_t(x)}}.
\end{equation}

\paragraph{Hyperbolic kinetic energy and nonlocal Fisher information.}
\label{subsubsec:hyperbolic-kinetic-functional}

\begin{definition}[Hyperbolic kinetic energy and nonlocal Fisher information]
\label{def:hamiltonian}
For a symmetric mobility $\theta$, define the hyperbolic kinetic energy
\begin{equation*}
\mathcal K_{\mathrm{hyper}}(\rho,\phi)
:=\iint_{M\times M}
\left[\cosh\bigl(\Nabla\phi(x,y)\bigr)-1\right]
\,J_\rho(\mathrm dx,\mathrm dy),
\end{equation*}
and the nonlocal Fisher information
\begin{equation*}
\mathcal I(\rho)
:=\frac12\mathcal A_K(\sqrt\rho,\sqrt\rho)
=\frac12\iint_{M\times M}|\Nabla\sqrt\rho|^2\,K(\mathrm dx,\mathrm dy).
\end{equation*}
\end{definition}
The square-root Dirichlet functional $\mathcal I$ is distinct from the
entropy-gradient norm $\mathscr F_{\mathcal E}$ in
Section~\ref{subsubsec:gradient-Hamiltonian-lift}. For geometric-mean mobility, $\mathcal K_{\mathrm{hyper}}$ has the
$\cosh$ structure of \cite[\S1.1, (1.8), (1.17a)]{Peletier+Rossi+Savare+Tse_2022_Jump}:
in the common setting, with the same edge measure,
$\mathscr R^*(\rho,2\Nabla\phi)=2\mathcal K_{\mathrm{hyper}}(\rho,\phi)$.
Here it enters the canonical equations through the Hopf--Cole transformation.

\begin{proposition}[Hamiltonian formulation of the Schr\"odinger system]
\label{prop:bridge-Hamiltonian}
For the geometric-mean mobility, \eqref{eq:continuity}--\eqref{eq:hj_raw}
are the canonical equations for
\begin{equation}
\label{eq:bridge-Hamiltonian}
\mathcal H_1(\rho,\phi)
=\mathcal K_{\mathrm{hyper}}(\rho,\phi)-\mathcal I(\rho)-\mathcal V(\rho).
\end{equation}
The Hopf--Cole representative fixes $C(t)=0$.
\end{proposition}

\begin{proof}
Nonlocal integration by parts gives
\begin{align*}
D_\phi\mathcal K_{\mathrm{hyper}}(\rho,\phi)[\psi]=\iint\sinh(\Nabla\phi)\Nabla\psi\,J_\rho(\mathrm dx,\mathrm dy)=-\int_M\psi\,\Div\bigl(J_\rho\sinh(\Nabla\phi)\bigr)\,\mathrm dx.
\end{align*}
For a density variation $\sigma$, symmetry of $\theta$ and $K$
allows the two mobility-variation terms to be combined:
\[
D_\rho\mathcal K_{\mathrm{hyper}}(\rho,\phi)[\sigma]
=2\iint\sigma(x)\partial_a\theta\bigl(\rho(x),\rho(y)\bigr)
\bigl[\cosh(\Nabla\phi)-1\bigr]\,K(\mathrm dx,\mathrm dy).
\]
For the geometric mean,
$2\partial_a\theta(a,b)=\sqrt{b/a}$, so
\begin{equation*}
\begin{aligned}
\frac{\delta\mathcal K_{\mathrm{hyper}}}{\delta\phi}(\rho,\phi)
&=-\Div\bigl(J_\rho\sinh(\Nabla\phi)\bigr),\\
\frac{\delta\mathcal K_{\mathrm{hyper}}}{\delta\rho}(\rho,\phi)(x)
&=\int_M\sqrt{\frac{\rho(y)}{\rho(x)}}
\bigl[\cosh\bigl(\Nabla\phi(x,y)\bigr)-1\bigr]K(x,y)\,\mathrm dy.
\end{aligned}
\end{equation*}
For $\mathcal I$, the nonlocal Green's formula yields
\[
D\mathcal I(\rho)[\sigma]
=\mathcal A_K\left(\sqrt\rho,\frac{\sigma}{2\sqrt\rho}\right)
=-\int_M\frac{\sigma}{2\sqrt\rho}L_K\sqrt\rho\,\mathrm dx.
\]
Hence
\begin{equation}
\label{eq:fisher_var}
\frac{\delta\mathcal I}{\delta\rho}(\rho)(x)
=-\frac{L_K\sqrt\rho(x)}{2\sqrt{\rho(x)}}\\
=-\mathrm{p.v.}\int_M\left[\sqrt{\frac{\rho(y)}{\rho(x)}}-1\right]K(x,y)\,\mathrm dy.
\end{equation}
Substituting these derivatives into
\eqref{eq:canonical-Hamiltonian-system} for
\eqref{eq:bridge-Hamiltonian} gives \eqref{eq:continuity} and
\begin{equation*}
\partial_t\phi_t+\frac{\delta\mathcal K_{\mathrm{hyper}}}{\delta\rho}(\rho_t,\phi_t)
=V+\frac{\delta\mathcal I}{\delta\rho}(\rho_t).
\end{equation*}
The latter is exactly \eqref{eq:hj_split}, so
Proposition~\ref{prop:bridge-Cole-Hopf} identifies this Hamiltonian
system with \eqref{eq:master}.
\end{proof}

\paragraph{Quadratic momentum structure.}

Differentiation in the potential variable gives
\begin{align*}
D_\phi\mathcal K_{\mathrm{hyper}}(\rho,\phi)[\psi]
&=\iint_{M\times M}
\sinh(\Nabla\phi)\Nabla\psi\,J_\rho(\mathrm dx,\mathrm dy),\\
D_\phi^2\mathcal K_{\mathrm{hyper}}(\rho,\phi)[\psi,\xi]
&=\iint_{M\times M}
\cosh(\Nabla\phi)\Nabla\psi\Nabla\xi\,J_\rho(\mathrm dx,\mathrm dy).
\end{align*}
At zero momentum, $\sinh0=0$ and $\cosh0=1$, hence
\begin{equation*}
D_\phi\mathcal K_{\mathrm{hyper}}(\rho,0)[\psi]=0,\quad
D_\phi^2\mathcal K_{\mathrm{hyper}}(\rho,0)[\psi,\xi]
=g_\rho(\Nabla\psi,\Nabla\xi).
\end{equation*}
Thus the momentum Hessian recovers the quadratic kinetic energy:
\begin{equation*}
\mathcal K(\rho,\phi)
=\tfrac12D_\phi^2\mathcal K_{\mathrm{hyper}}(\rho,0)[\phi,\phi].
\end{equation*}

Using the full power series of $\cosh$
gives the following identity
\begin{equation*}
\begin{aligned}
\mathcal K_{\mathrm{hyper}}(\rho,\epsilon\phi)
&=\sum_{n=1}^{\infty}\frac{\epsilon^{2n}}{(2n)!}
\iint_{M\times M}|\Nabla\phi|^{2n}\,J_\rho(\mathrm dx,\mathrm dy)\\
&=\epsilon^2\mathcal K(\rho,\phi)
+\sum_{n=2}^{\infty}\frac{\epsilon^{2n}}{(2n)!}
\iint_{M\times M}|\Nabla\phi|^{2n}\,J_\rho(\mathrm dx,\mathrm dy).
\end{aligned}
\end{equation*}
Here the $n=1$ term is $\epsilon^2\mathcal K(\rho,\phi)$ by
\eqref{eq:quadratic-kinetic-energy}.
For a smooth finite-energy $\phi$ on the torus and
$0<|\epsilon|\leq\epsilon_0$, Taylor's formula gives
\[
0\leq\frac{\cosh(\epsilon\Nabla\phi)-1}{\epsilon^2}
\leq\frac12\cosh(2\epsilon_0\|\phi\|_{L^\infty(M)})|\Nabla\phi|^2.
\]
The right-hand side is integrable against $J_\rho$, so dominated
convergence yields
\begin{equation*}
\lim_{\epsilon\to0}\epsilon^{-2}\mathcal K_{\mathrm{hyper}}(\rho,\epsilon\phi)
=\mathcal K(\rho,\phi).
\end{equation*}
The momentum Hessian at zero therefore recovers the metric $g_\rho$,
whereas the full Hamiltonian is nonquadratic.

\subsubsection{Nonlocal Schr\"odinger equation}
\label{subsec:nonlocal-Schrodinger-Madelung-revised}

Use the symmetric generator $\frac12L_K$ from
\eqref{eq:examples-LcalK} and the geometric-mean mobility
$J_\rho(x,y)=\sqrt{\rho(x)\rho(y)}K(x,y)$.
In units with the phase parameter equal to $1$, consider
\begin{equation}
\label{eq:examples-complex-Schrodinger}
\mathrm i\,\partial_t\Psi_t(x)
=-\frac12L_K\Psi_t(x)+V(x)\Psi_t(x),
\end{equation}
where $V$ is real valued and $\Psi_t:M\to\mathbb C$.
For $K=K_s$ on $\mathbb R^d$ or its periodized counterpart on
$\mathbb T^d$, the kinetic operator is $(-\Delta)^s$.
We normalize $\int_M|\Psi_0|^2\,\mathrm dx=1$; symmetry of
$\frac12L_K$ and reality of $V$ preserve this mass.

For a nonvanishing wave function with a globally defined real phase,
the \emph{Madelung transform} is
\begin{equation}
\label{eq:examples-Madelung-transform}
\Psi_t(x)=\sqrt{\rho_t(x)}\,e^{\mathrm i\phi_t(x)},
\qquad \rho_t(x)>0.
\end{equation}

\begin{proposition}[Madelung and Hamiltonian formulations]
\label{prop:nonlocal-Madelung-revised}
Under \eqref{eq:examples-Madelung-transform}, the smooth nonvanishing
solutions of \eqref{eq:examples-complex-Schrodinger} correspond to
smooth pairs $(\rho_t,\phi_t)$ satisfying
\begin{equation}
\label{eq:examples-quantum-continuity}
\partial_t\rho_t+\Div\bigl(J_{\rho_t}\sin(\Nabla\phi_t)\bigr)=0
\end{equation}
and
\begin{equation}
\label{eq:examples-quantum-HJ}
\partial_t\phi_t(x)
=\mathrm{p.v.}\int_M\left[
\sqrt{\frac{\rho_t(y)}{\rho_t(x)}}
\cos\bigl(\Nabla\phi_t(x,y)\bigr)-1
\right]K(x,y)\,\mathrm dy-V(x).
\end{equation}
These are the canonical equations generated by the Hamiltonian
\begin{equation}
\label{eq:examples-Madelung-energy}
\mathcal H_2(\rho,\phi)
=\mathcal I(\rho)
+\iint_{M\times M}\bigl[1-\cos(\Nabla\phi)\bigr]
\,J_\rho(\mathrm dx,\mathrm dy)
+\mathcal V(\rho).
\end{equation}
The phase representative fixes $C(t)=0$; $\phi_t$ is the conjugate
potential, not in general the metric velocity potential of $\rho_t$.
\end{proposition}

\begin{proof}
In contrast to the real exponentials in
Proposition~\ref{prop:bridge-Cole-Hopf}, the complex phase gives
\[
\begin{aligned}
\frac{\mathrm i}{2\rho_t(x)}\partial_t\rho_t(x)-\partial_t\phi_t(x)
&=-\mathrm{p.v.}\int_M\left[
\sqrt{\frac{\rho_t(y)}{\rho_t(x)}}
e^{\mathrm i\Nabla\phi_t(x,y)}-1
\right]K(x,y)\,\mathrm dy+V(x).
\end{aligned}
\]
Taking imaginary and real parts yields
\eqref{eq:examples-quantum-continuity} and
\eqref{eq:examples-quantum-HJ}. Conversely, reversing this calculation
reconstructs the Schr\"odinger equation.

The identity
\[
|\Psi(y)-\Psi(x)|^2
=\bigl(\sqrt{\rho(y)}-\sqrt{\rho(x)}\bigr)^2+2\sqrt{\rho(x)\rho(y)}
\bigl[1-\cos\bigl(\Nabla\phi(x,y)\bigr)\bigr]
\]
proves \eqref{eq:examples-Madelung-energy}.
The integration-by-parts and mobility-variation calculations in
Proposition~\ref{prop:bridge-Hamiltonian} now use the phase energy
$1-\cos(\Nabla\phi)$ and the positive amplitude energy $\mathcal I$.
They give
\[
\begin{aligned}
\frac{\delta\mathcal H_2}{\delta\phi}(\rho,\phi)
&=-\Div\bigl(J_\rho\sin(\Nabla\phi)\bigr),\\
\frac{\delta\mathcal H_2}{\delta\rho}(\rho,\phi)(x)
&=V(x)+\frac{\delta\mathcal I}{\delta\rho}(\rho)(x)+\int_M\sqrt{\frac{\rho(y)}{\rho(x)}}
\bigl[1-\cos\bigl(\Nabla\phi(x,y)\bigr)\bigr]K(x,y)\,\mathrm dy\\
&=V(x)+\mathrm{p.v.}\int_M\left[
1-\sqrt{\frac{\rho(y)}{\rho(x)}}\cos\bigl(\Nabla\phi(x,y)\bigr)
\right]K(x,y)\,\mathrm dy,
\end{aligned}
\]
Since $0\leq 1-\cos z\leq z^2/2$, the phase integral in the first
expression for $\delta\mathcal H_2/\delta\rho$ converges absolutely for almost every
$x$. The principal value in the combined expression comes from the amplitude
derivative \eqref{eq:fisher_var}.
Substitution into \eqref{eq:canonical-Hamiltonian-system} proves the
Hamiltonian assertion.
\end{proof}




\section*{Statements and declarations}

\paragraph{Funding.}
The work of Y.~Chen and J.~Duan was partly supported by the NSFC  International Collaboration Fund for Creative Research Teams (Grant W2541005), the Guangdong Provincial Key Laboratory of Mathematical and Neural Dynamical Systems (Grant 2024B1212010004), the Cross Disciplinary Research Team on Data Science and Intelligent Medicine (2023KCXTD054), the Guangdong-Dongguan Joint Research Fund (Grant 2023A151514 0016), and the Dongguan Key Laboratory for Data Science and Intelligent Medicine.
The work of Q.~Huang was supported by the National Natural Science Foundation of China
under Grant No.~12501241, the Basic Research Program of Jiangsu Province under Grant
No.~BK20251280, the Fundamental Research Funds for the Central Universities
under Grant No.~2242026RCB0008 (Zhishan Young Scholars), the Jiangsu Provincial Scientific Research
Center of Applied Mathematics under Grant No.~BK20233002, and the Start-up
Research Fund of Southeast University under Grant No.~RF1028624194.

\paragraph{Data availability.} We do not analyze or generate any datasets, because our work proceeds within a theoretical and mathematical approach.

\paragraph{Competing interests.} The authors have no competing interests to declare that are relevant to the content of this article.

\paragraph{Declaration of generative AI.} OpenAI ChatGPT was used to assist with proof
auditing and grammar corrections. The authors reviewed and edited the content thoroughly as needed and take full responsibility for the content of the published article.

\bibliographystyle{abbrv}
\bibliography{Refs}

\end{document}